\documentclass[10pt,showpacs,secnumarabic,nofootinbib,onecolumn,aps,prx,longbibliography,superscriptaddress,notitlepage ]{revtex4-2}
\usepackage[utf8]{inputenc}
\usepackage{amsmath}
\usepackage{amsthm}
\usepackage{amsfonts}
\usepackage{amssymb}
\usepackage{amssymb}
\usepackage{bm}
\usepackage{times, txfonts}
\usepackage[colorlinks,citecolor=blue,urlcolor=blue,linkcolor=blue]{hyperref}
\usepackage{booktabs}
\usepackage{algorithm}
\usepackage{algpseudocode}
\usepackage[capitalise]{cleveref}
\usepackage{physics}
\usepackage{graphicx}
\usepackage{subcaption}
\usepackage{color}
\usepackage[dvipsnames]{xcolor}
\usepackage{url}
\usepackage{bbm}
\usepackage{rotating}
\usepackage[normalem]{ulem}
\usepackage{orcidlink}
\usepackage{pifont}

\crefname{supp}{supplementary material}{supplementary materials}
\Crefname{supp}{Supplementary Material}{Supplementary Materials}

\theoremstyle{definition}

\theoremstyle{plain}

\theoremstyle{remark}

\makeatletter
\newcommand*\bigcdot{\mathpalette\bigcdot@{.5}}
\newcommand*\bigcdot@[2]{\mathbin{\vcenter{\hbox{\scalebox{#2}{$\m@th#1\bullet$}}}}}
\makeatother
\newcommand{\ii}{\mathrm{i}}
\newcommand{\mc}{\mathcal}
\newcommand{\bv}{\mathbf}

\newcommand{\rev}[1]{{#1}}
\newcommand{\Tsinghua}{Department of Chemistry and Engineering Research Center of Advanced Rare-Earth Materials of Ministry of Education, Tsinghua University, Beijing 100084, China}

\newcommand{\SUSTECH}{Department of Chemistry and Guangdong Provincial Key Laboratory of Catalytic Chemistry, Southern University of Science and Technology, Shenzhen 518055, China}
\newcommand{\GZ}{Fundamental Science Center of Rare Earths, Ganjiang Innovation Academy, Chinese Academy of Sciences, Ganzhou 341000, China}
\begin{document}
%\input{preamble.tex}
%\preprint{APS/123-QED}

\title{Towards robust variational quantum simulation of Lindblad dynamics \\
via stochastic Magnus expansion}% Force line breaks with \\

%\author{Authors}
%\affiliation{Department}

\author{Jia-Cheng Huang\orcidlink{0009-0001-8245-1432}}
\thanks{These authors contributed equally to this work.}

\affiliation{\Tsinghua}%

\author{Hao-En Li\orcidlink{0009-0002-2807-2826}}
\thanks{These authors contributed equally to this work.}

\affiliation{\Tsinghua}%
\affiliation{Department of Mathematics, University of California, Berkeley, California 94720, USA}
\author{Yi-Cheng Wang\orcidlink{0009-0004-4641-0782}}
\affiliation{\Tsinghua}%

\author{Guang-Ze Zhang\orcidlink{0009-0007-6272-5831}}
\affiliation{\Tsinghua}%

\author{Jun Li\orcidlink{0000-0002-8456-3980}}
\affiliation{\Tsinghua}%
\affiliation{\SUSTECH}
\affiliation{\GZ}

\author{Han-Shi Hu\orcidlink{0000-0001-9508-1920}}
\email{hshu@mail.tsinghua.edu.cn}
\affiliation{\Tsinghua}%

\date{\today}

\begin{abstract}
\centerline{\textit{Dedicated to the centenary of quantum mechanics.}}

 In this paper, we introduce a novel and general framework for the variational quantum simulation of Lindblad equations. Building on the close relationship between the unraveled Lindblad dynamics, stochastic Magnus integrators, and variational quantum simulation, we propose a high-order scheme for solving the quantum state diffusion equation using exponential integrators. This formulation facilitates the simulation of wavefunction trajectories within the established framework of variational quantum algorithms for time evolution. Our algorithm significantly enhances robustness in two key aspects: the stability of the simulation with large time steps, and the reduction in the number of quantum trajectories required to accurately simulate the Lindblad dynamics in terms of the ensemble average. We demonstrate the effectiveness of our algorithm through numerical examples in both classical and quantum implementations, including the transverse-field Ising model (TFIM) with damping, the Fenna–Matthews–Olson (FMO) complex, and the radical pair model (RPM). The simulation accuracy can be systematically improved, and the algorithm remains reliable even in highly oscillatory regimes. These methods are expected to be applicable to a broader class of open quantum systems beyond the specific models considered in this study. 
\end{abstract}

\maketitle

%\tableofcontents

%\newpage
\section{Introduction}

The dynamics of systems governed by Lindblad equations (or Lindblad master equations, LMEs) \cite{Lindblad1976,  GoriniKossakowskiSudarshan1976,Davies1974} give rise to a range of unique physical phenomena, which have attracted considerable theoretical and experimental interest in recent studies. In contrast to unitary dynamics described by the Schrödinger equation, Lindblad dynamics accounts for the interaction between the system and its environment through the introduction of a set of Lindblad jump operators (or bath operators). Due to its versatile modeling capability, Lindblad dynamics can describe a wide range of non-Hermitian quantum behaviors arising in diverse fields. Notably, it can be employed to model processes such as light-harvesting in photosynthesis \cite{PalmieriAbramaviciusMukamel2009}, charge and energy transfer in quantum dots and molecules \cite{HarbolaEspositoMukamel2006}, and radical pair dynamics in magnetic resonance \cite{PooniaKondabagilSahaGanguly2017,AdamsSinayskiyPetruccione2018}.
In addition to its applicability for simulating realistic system-bath interactions, recent advancements have shown that Lindblad dynamics can be a powerful tool for approximation of non-equilibrium dynamics \cite{MascherpaSmirneHuelgaPlenio2017, TamascelliSmirneHuelgaPlenio2018}, quantum error correction \cite{Terhal2015, LaydenZhouCappellaroJiang2019}, and efficient preparation of quantum Gibbs states and ground states on quantum computers \cite{ChenKastoryanoGilyén2023, ChenKastoryanoBrandãoEtAl2023, DingChenLin2024, LiZhanLin2024}. These developments have generated significant interest across the fields of quantum information, quantum algorithms, quantum many-body physics, and quantum chemistry.

Given the extensive and significant applications of the Lindblad equation, we aim to develop robust numerical methods for simulating Lindblad dynamics, which have the potential to benefit various fields. However, since the many-body density operator has the dimension that scales exponentially with the system size, it can be prohibitively expensive to solve the Lindblad equation using classical numerical methods \cite{BreuerPetruccione2002,BieleD’Agosta2012, CaoLu2021, ChenLiLuYing2024}. From the perspective of quantum computing, Lindblad evolution describes a completely positive and trace-preserving (CPTP) map, also known as a quantum channel, potentially enabling efficient simulation on quantum computers. While theoretical quantum algorithms that attain near-optimal complexity have been developed \cite{ChildsLi2017, CleveWang2017, LiWang2023, DingLiLin2024}, their implementation is typically much more complicated than that of Hamiltonian simulation \cite{BerryAhokasCleveSanders2007, LowChuang2019}, mainly owing to the non-Hermitian nature of the problem \cite{DingLiLin2024}. Furthermore, there has been a limited investigation into the potential for developing robust \emph{near-term} quantum algorithms that can effectively simulate Lindblad dynamics \cite{DelgadoGranadosLuisKrogmeierEtAl2024}%\HL{ASAP, update it when the issue and volume numbers are available}
. 

\begin{figure}[htbp]
    \centering
    \includegraphics[width=0.98\linewidth]{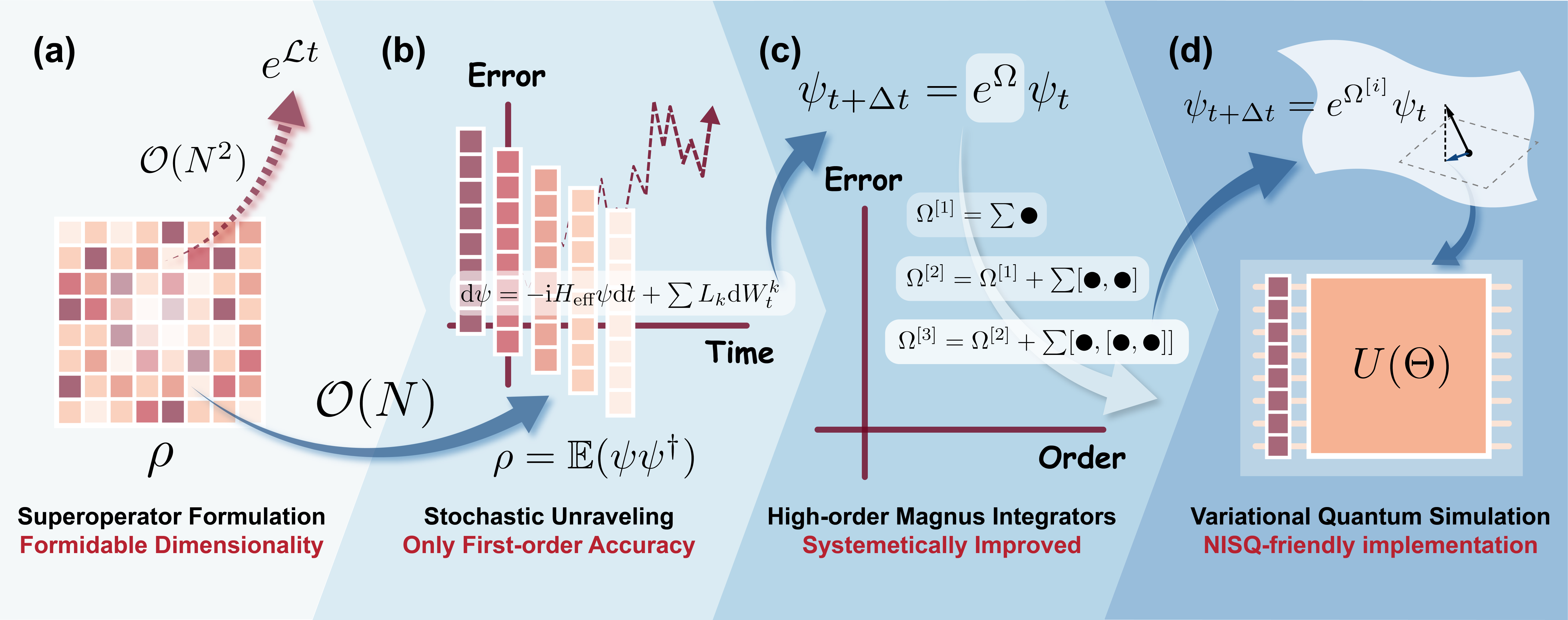}
    \caption{\raggedright Conceptual illustration of our algorithm and its advantages. (a) Solving the Lindblad equation with the superoperator formulation is faced with a formidable dimensionality of $4^L$ (the size of the many-body density operator). (b) Unraveled Lindblad dynamics (e.g. QSD) offers a Monte Carlo trajectory-based method that reduces dimensionality quadratically, but its original version has only first-order accuracy and suffers from sampling variance. (c) High-order stochastic Magnus expansions systematically improve precision and robustness. (d) The form of exponential integrators makes them suitable for variational quantum simulation. }
    \label{fig:illus}
\end{figure}

Existing near-term approaches include techniques such as unitary decomposition of operators \cite{SchlimgenHeadMarsdenSagerEtAl2021, SchlimgenHeadMarsdenSagerEtAl2022}, 
Sz.-Nagy unitary dilation \cite{HuXiaKais2020, HuHeadMarsdenMazziottiEtAl2022, ZhangHuWangKais2023}, hybrid quantum channel approach \cite{HannLeeGirvinJiang2021,PeetzSmartTserkisNarang2024}, quantum imaginary time evolution (QITE) \cite{KamakariSunMottaMinnich2022}, singular value decomposition-based quantum simulation \cite{SchlimgenHeadMarsdenSagerSmithEtAl2022,OhKrogmeierSchlimgenEtAl2024}, and variational quantum simulation (VQS) of quantum dynamics \cite{ChenGomesNiuEtAl2024, LuoLinGao2024, LanLiang2024}. 
 Among these, the first five algorithms correspond to either vectorized superoperator formulation or Kraus-form formulation of the Lindblad equation. These two formulations both operate within a Hilbert space of dimensionality $4^L$, where $L$ is the number of qubits, which corresponds to the size of the many-body density operator (\cref{fig:illus} 
 (a)). As a result, these methods are computationally expensive and less suitable for near-term quantum devices. In contrast, variational simulation methods, which only require simulating a wavefunction of dimension $2^L$, are more efficient in this context.

The central concept of the variational simulation algorithm for Lindblad dynamics is to \emph{unravel} the Lindblad equation and reformulate it as a stochastic differential equation (SDE) for the wavefunction \cite{BreuerPetruccione2002,GisinPercival1992,Percival1998,Lidar2020}. This formulation primarily includes two algorithms: quantum state diffusion (QSD) and quantum jump (QJ). Following this reformulation, we obtain a Monte Carlo-type wavefunction trajectory algorithm for numerical solution of the Lindblad equation. This idea can also be integrated with variational quantum algorithms to solve the stochastic non-Hermitian dynamics for multiple wavefunction trajectories, and the ensemble average of these pure-state trajectories will converge to the exact many-body density operator. However, due to the variance of the samplings of the random processes, many existing algorithms based on unraveled Lindblad dynamics require a large number of wavefunction samples, extremely small time steps, or other hyperparameters that need careful tuning to achieve a high simulation fidelity. As a result, these algorithms often lack robustness, particularly when high accuracy is required for simulating the Lindblad equation (\cref{fig:illus} (b)) \cite{ChenGomesNiuEtAl2024,LuoLinGao2024, LanLiang2024}. 

In this paper, we introduce a novel implementation of the QSD-based variational simulation. Our algorithm exploits the deep connection among McLachlan’s variational principle for quantum dynamics, the stochastic Magnus expansion for the QSD equation, and exponential integrators for solving SDEs. Specifically, as illustrated in \cref{fig:illus} (c), we derive a new high-order numerical scheme for the linear and nonlinear QSD equations using the stochastic Magnus expansion. A hallmark of such numerical scheme is that it takes the form of an exponential integrator, making the variational simulation feasible (\cref{fig:illus} (d)). By incorporating numerical schemes of arbitrarily high order, our approach offers a systematic means to enhance the robustness of variational quantum simulation compared to other variational methods, both in terms of ensemble averages and time step management. This enables accurate and stable simulation results at a relatively low cost, while preserving the qubit efficiency inherent to the variational algorithm. Furthermore, the form of Magnus expansion can be explicitly characterized through the evaluation of multiple stochastic integrals, calculated via the Fourier expansion of Brownian bridges, which facilitates the easy derivation of higher-order precision schemes, and the time evolution based on the variational algorithm can be naturally extended to time-dependent scenarios.

The remainder of this paper is organized as follows. In \cref{sec:prelim}, we provide some preliminaries on QSD methods, introduce key notations, and explain the connection between the SDEs and the Lindblad equation. Additionally, we outline the framework of VQS for non-Hermitian quantum dynamics, and clarify its relationship with exponential integrators. 
\Cref{sec:stomagnus} presents the fundamental concepts of stochastic Magnus integrators and details our main approach for constructing high-order numerical schemes to solve the QSD equation. The effectiveness of the proposed algorithm is demonstrated through a series of numerical experiments in \cref{sec:numerics}. %In \cref{sec:VQS}, we offer a brief overview of the variational quantum simulation framework and clarify its connection to exponential integrators. 
Additionally, \cref{appA} provides a detailed comparison between linear and nonlinear unraveling. \cref{app:magnus_taylor} provides some insights on the conceptual difference between the stochastic Magnus expansion and other approaches for deriving high-order schemes for SDEs. The technical derivations supporting our numerical schemes can be found in Appendices \ref{appB} and \ref{appC}. An illustrative derivation of the Magnus integrators for each order is given in \cref{appD}. The implementation details of our numerical tests and further numerical results 
 are presented in \cref{appE,app:EM_magnus_compare,appF}.

\section{Preliminaries}\label{sec:prelim}

Throughout the paper, we use $\ket{\psi}$ for ($\ell^2$-) normalized state vectors and $\psi$ for unnormalized state vectors in quantum dynamics. For a vector $\psi \in \mathbb C^n$, $\abs{\psi}$ denotes the $\ell^2$-norm. For a matrix $A\in \mathbb C^{n\times n}$, $A^\ast$ and $A^\dagger$ denote the complex conjugation and conjugate transpose (or adjoint) of $A$ respectively. $\norm{A}$ is the operator $2$-norm (or spectral norm) of $A$. We use $\psi^i$ and $A^{i,j}$ to denote the $i$-th entry of the vector $\psi$ and the $(i,j)$-entry of the matrix $A$ respectively when the context is clear.
\subsection{Quantum state diffusion unraveling of Lindblad equation}\label{sec:unravel}

With the Born-Markov and rotating wave approximations \cite{BreuerPetruccione2002}, the dynamics of an open quantum system can be described by the Lindblad master equation (LME)
\begin{equation} \label{lindblad}
\dv{\rho}{t} = \mathcal{L}[\rho] = -{\ii}[{H}_s,\rho] + \sum_{k} {L}_k\rho{L}_k^\dagger - \frac{1}{2}\sum_{k} \{{L}_k^\dagger {L}_k,\rho\},
\end{equation}
where $\rho$ is the system many-body density operator, ${H}_s$ is the system Hamiltonian, $\{{L}_k\}$ are bath operators, and $\{a,b\}=ab+ba$ denotes the anticommutator. For simplicity, here the dissipation rate $\Gamma_k$ has been absorbed in bath operator like $\sqrt{\Gamma_k}L_k\mapsto L_k$. Directly propagating the density operator $\rho$ under \cref{lindblad} encounters substantial difficulties due to the intricate structure of the superoperator (also known as Lindbladian) $\mathcal{L}$, as well as the $4^n$-exponential scaling of the underlying Hilbert space dimension. An alternative approach is based on the stochastic unraveling of LME, in which the density operator is represented by the ensemble mean over pure state projection operators,
\begin{equation}
    \rho = \mathbb E(\psi\psi^\dagger).
\end{equation}

There are various types of pure state dynamics for $\psi$ that are consistent with the LME in the sense of ensemble averaging, which means that $\mathbb E(\psi \psi^\dagger)$ could reproduce the Lindblad dynamics \cref{lindblad} as
\begin{equation}\label{eq:reproduceLME}
    \frac{\dd \mathbb E (\psi\psi^\dagger)}{\dd t} = -{\ii}[{H}_s,\mathbb E (\psi\psi^\dagger)] + \sum_{k} {L}_k(\mathbb E (\psi\psi^\dagger)){L}_k^\dagger - \frac{1}{2}\sum_{k} \{{L}_k^\dagger {L}_k,\mathbb E (\psi\psi^\dagger)\}.
\end{equation}
For instance, the QJ unraveling, as introduced by \citet{Carmichael1993}, corresponds to propagating the wave function trajectories according to an SDE driven by Poisson process. In this paper, the QSD approach governed by continuous stochastic process is employed, where $\psi$ satisfies the following Itô-type SDE:
\begin{equation}\label{eq:nlqsd}
\mathrm{d} \psi=  -{\ii} {H}_s\psi \mathrm{d} t+\sum_k(\langle{L}_k^{\dagger}\rangle {L}_k-\frac{1}{2} {L}_k^{\dagger} {L}_k-\frac{1}{2}\langle{L}_k^{\dagger}\rangle\langle{L}_k\rangle)\psi \mathrm{d} t  +\sum_k({L}_k-\langle{L}_k\rangle)\psi \mathrm{d} W_t^{k}.
\end{equation}
Here, $\langle\cdot\rangle \equiv \psi^\dagger (\cdot)\psi$ denotes the (unnormalized) expectation value of the corresponding operator, and stochastic fluctuation $\{\mathrm{d} W_t^{k}\}$ are independent Wiener (or Gaussian) processes satisfying the conditions

\begin{equation}
    \mathbb E(\mathrm{d}W_t^k)=0, \quad \mathbb E(|\mathrm{d}W_t^k|^2)=\mathrm{d}t,\quad \mathbb E(\mathrm{d}W_t^{{k}^*}\mathrm{d}W_t^j)=0\, (k\neq j).
\end{equation}
In fact, within the scope of QSD, we can omit the nonlinear term $\langle{L}_k^{\dagger}\rangle$ and $\langle L_k\rangle$ to obtain a much simpler linear QSD scheme, as has been used in many existing works~\cite{LiLi2019,DingLiLin2024,ChenGomesNiuEtAl2024, LuoLinGao2024}:
\begin{equation}\label{eq:linear}
    \dd \psi = -\ii  H_s\psi \dd t -\frac12\sum_k L_k^\dagger L_k \psi\dd t + \sum_k L_k\psi \dd W_t^k = -\ii H_{\rm eff} \psi \dd t +\sum_k L_k \psi \dd W_t^k.
\end{equation}
Here $H_{\rm eff}$ is the non-Hermitian effective system Hamiltonian defined as $H_{\rm eff} = H_s -\frac\ii 2\sum_k L_k^\dagger L_k$. This notation will also be used later in this paper. Using Itô’s formula for $\psi\psi^\dagger$ and taking the expectation, one can verify that both nonlinear and linear QSD unraveling satisfy \cref{eq:reproduceLME}. 

The linear SDE \cref{eq:linear} is more concise in form. However, when implementing time discretization to develop a Monte Carlo trajectory-based algorithm, it can be proved that the norm of individual wave function $\psi$ is \emph{not} preserved up to $\mathcal{O}(\Delta)$ in the linear unraveling. This leads to an ensemble weighted by the changing norm, inducing large fluctuations and increasing variance. Despite the fact that the variance of the ensemble average of the trajectories remains independent of the specific unraveling method prior to time discretization, this property is generally \emph{not} preserved after discretization \cite{LeBrisRouchonRoussel2015}. While there are effective variance-reduction techniques based on adaptive low-rank approximations, these methods are relatively complex and less suited for implementation on quantum devices \cite{LeBrisRouchonRoussel2015}. On the contrary, we adopt a more straightforward approach by turning to nonlinear formulation in some cases. Although the nonlinear QSD \cref{eq:nlqsd} seems more complex at first glance, the \emph{norm-preserving} property of the nonlinear scheme ensures that the ensemble is not dominated by samples with large norms, preventing the variance from growing excessively at discrete time steps. For a more detailed discussion, we refer the reader to \cref{appA}.

\subsection{Sketch of numerical schemes for SDEs}\label{sec:sketch_numerSDE}

After reformulating the Lindblad dynamics, solving SDEs \cref{eq:nlqsd} or \cref{eq:linear} becomes the central component of the algorithm. In this paper, we consider the stochastic Magnus expansion for robust numerical simulation of the unraveled dynamics. 

In this section, we will give a brief introduction to some well-developed numerical schemes for solving SDEs. Here we only discuss the explicit numerical scheme, since it has a more direct implementation in quantum computing. It is worth noting that both the nonlinear QSD \cref{eq:nlqsd} and the linear QSD \cref{eq:linear} are autonomous SDEs, meaning that the equations do not explicitly depend on the time variable and only involve the wavefunction $\psi$. In general, we consider the following general autonomous Itô-type SDE:
\begin{equation}
    \dd X_t=a(X_t)\dd t + b(X_t) \dd W_t.
\end{equation}
Comparing with the ordinary differential equations (ODEs), there are subtleties in treating stochastic models owing to the Wiener increments $\dd W_t$. %and not many numerical methods are well-developed. 
As in the deterministic case, the numerical simulation of the SDE also requires discretizing total time $T$ into finite steps $N =T/\Delta~$\cite{Gillespie1996, KloedenPlaten2013}. There are different convergence criteria for the time-discretization approximation, including weak and strong convergence. For a stochastic process $X_t$, the corresponding time-discretization approximation $Y_k(k=0,1,\dots,N)$ is said to converge weakly with order $\beta\in (0,\infty]$ if there exists a finite constant $C$ and a positive constant $t_0$ such that
\begin{equation}
    \abs{\mathbb E(g(X_T))-\mathbb E(g(Y_N))}\leq C {\Delta}^\beta
\end{equation}
for any polynomial $g$ and $\Delta \in (0, t_0)$. On the other hand, the strong convergence requires the approximation trajectory be sufficiently close to the sample path such that
\begin{equation}
    \mathbb E(\abs{X_T-Y_N})\leq C{\Delta }^\alpha,
\end{equation}
where $\alpha$ denotes the strong convergence order, and the rest is the same as above. To establish a numerical scheme with ideal convergence order, a natural approach is to extend the deterministic single-step methods, commonly used for ODEs, into the stochastic scenarios. The simplest scheme among them is the Euler-Maruyama method, resembling the explicit Euler method for deterministic ODEs. Specifically, we take the linearly unraveled Lindblad equation \cref{eq:linear} as an example, then the corresponding Euler-Maruyama scheme is given by
\begin{equation}\label{eq:EulerMaruyama}
    \psi_{n+1} = \psi_n - \ii H_{\rm eff} \psi_n \Delta  + \sum_{k} L_k\psi_n\sqrt{\Delta } W^k.
\end{equation}
Here, $\{W^k\}_k$ are independent standard Gaussian random variables (with zero mean and unit variance), and $\sqrt{\Delta } W^k$ can be viewed as the time discretization of $\dd W_t^k$. This numerical scheme can provide the solution with a weak convergence of order $\mathcal{O}(\Delta)$ \cite{Burrage1999, BurrageBurrage1999,DingLiLin2024}, thus leading to
\begin{equation}
  \norm{  \mathbb E(\psi_{n+1}\psi_{n+1}^\dagger) - \mathbb E (\psi_n \psi_n^\dagger)} = \mc O(\Delta ).
\end{equation}
However, it has been proved that the methods using only Wiener increments can not give rise to strong convergence of order higher than one. Therefore, to achieve a robust and precise simulation of SDEs, higher order approximation which contains more information about Wiener process must be employed~\cite{BurrageBurragePamelaDesmond2006}. Such approximations can be obtained by truncating the stochasitc Taylor expansion.  For example,  the Milstein scheme, which contains one additional term in expansion, has correspondingly higher accuracy than Euler-Maruyama by half an order of strong convergence. 

Although the stochastic Taylor expansion provides a feasible method, it still faces the following issues:
\begin{itemize}
    \item The derivation of expansion is complex and not compact enough, especially for higher order expansion and multiple dimension SDEs;
    \item As in the classical Taylor expansion, it involves the derivatives of the coefficients $a$ and $b$, which is intolerable in nonlinear case;
    \item The inappropriate truncation may disrupt the geometric structures and dynamic properties in physical system. 
\end{itemize}
For detailed conceptual comparison between the stochastic Taylor expansion and the stochastic Magnus expansion, we refer interested readers to \cref{app:magnus_taylor}.

We note that, in the previous discussion, only single-step methods are considered. While typical multi-step methods, for example, the Runge-Kutta method in stochastic version, provide higher accuracy and better stability, they often come with tedious coefficients and inherit the above drawbacks of single-step methods. Therefore, in our algorithm, we use an alternative approach based on Lie group, referred to as the \emph{stochastic Magnus expansion}. In the deterministic case, the Magnus expansion has shown a more efficient convergence performance~\cite{GabrielSimonAnke2007}, and the truncated series very often shares important qualitative properties with the exact solution. To mitigate the errors induced by nonlinear terms in SDEs, we also employ a straightforward multi-step approximation based on Magnus expansion, avoiding the complicated derivations or tedious coefficient matrices in stochastic Runge-Kutta methods.  The details of the algorithm are provided in \cref{sec:stomagnus}.

\subsection{Variational quantum simulation algorithm}

Propagating Lindblad dynamics on quantum circuits entails many challenges, whether utilizing vectorized superoperator and Kraus-form formulation or employing stochastic unraveling. Many algorithms often require continuously extending the length of the quantum circuit during evolution~\cite{ChenGomesNiuEtAl2024, HuHeadMarsdenMazziottiEtAl2022}, which means that long-term simulations are not feasible on existing Noisy Intermediate-Scale Quantum (NISQ) computers through them. In this work, we employ variational quantum simulation (VQS) algorithm to evolve the single trajectory $|\psi\rangle$ in QSD unraveling on the quantum device. VQS combines classical optimization techniques with quantum computing to approximate the dynamics and properties of complex quantum systems, including imaginary time evolution and other non-Hermitian quantum dynamics~\cite{McArdleJonesEndoEtAl2019,YuanEndoZhaoEtAl2019,GaconNysRossiEtAl2024,LiLiHuangEtAl2024}. While the performance of VQS is dependent on the ansatz design and classical optimization algorithm, it is straightforward to implement on NISQ hardware compared to many other types of algorithms, as the quantum resources it requires are fixed during the entire simulation. Specifically, in VQS algorithm, a parametrized quantum circuit
\begin{equation}
|\psi(\Theta(t))\rangle=U(\Theta(t))|0\rangle^{\otimes N}, \quad \Theta (t) = (\theta_1(t),\theta_2(t),\dots,\theta_n(t))
\end{equation}
is used as a variational wave function ansatz in advance, where $U(\Theta(t))$ is the product of unitary operators corresponding to quantum gates. In this work, we focus on the Hamiltonian variational ansatz (HVA), which is a natural ansatz for solving quantum many-body Hamiltonian. It has been shown that the parameters in HVA can be constrained to avoid barren plateaus~\cite{ParkKilloran2024}. After decomposing a given Hamiltonian ${H}=\sum_{i=1}^n c_i {H}_i$, where $\{c_i\}$ are coefficients and $\{{H}_i\}$ are Pauli strings, the corresponding HVA can be represented as
 \begin{equation}
     |\psi(\Theta(t))\rangle = \prod_{p=1}^m \left ( e^{-\ii H_n \theta_{n,p}}e^{-\ii H_{n-1} \theta_{n-1,p}}\cdots e^{-\ii H_1 \theta_{1,p}}\right)|\psi_0\rangle.
 \end{equation}
Here, $m$ is the number of blocks, and $n$ is the number of layers in each block, leading to a total of $mn$ parameters in HVA. Notably, this variational ansatz based method is applicable for non-Hermitian quantum dynamics while does not rely on the tomography-based procedures or quantum Monte Carlo \cite{MottaSunTanEtAl2020, HugginsOGormanRubinEtAl2022}.  

In this work, we aim to propagate the wavefunction trajectory according to
\begin{equation}
     \psi_{n+1} = \exp(\Omega(\Delta ))\psi_n = \exp(-\ii \mathcal H \Delta ) \psi_n.
 \end{equation}
Here $\Omega(\Delta )$ is the Magnus integrator defined in \cref{sec:stomagnus}. $\mathcal H \equiv \ii \Omega(\Delta )/\Delta $ denotes the non-Hermitian effective Hamiltonian derived from the Magnus integrator $\Omega (\Delta )$. According to the theory of variational quantum simulation~\cite{YuanEndoZhaoEtAl2019}, the evolution of states in Hilbert space then can be projected onto the tangent space of the variational ansatz by
\begin{equation}\label{eq:mclachlan}
    \delta \abs{\left | \dv{ \psi(\Theta(t))}{t}\right\rangle + \ii {\mc H} |\psi(\Theta(t))\rangle }^2=0,
\end{equation}
which is the well-known McLachlan’s variational principle. The equation of motion (EOM) can be derived as 
\begin{equation}\label{eq:eom}
    \mathbf{M}(t)\dot{\Theta} (t) = \mathbf{V}(t), 
\end{equation}
where the matrix elements
\begin{equation}
    M_{i,j} = \text{Re}\left\langle \frac{\partial \psi(\Theta (t))}{\partial \theta _i}\Bigg | \frac{\partial \psi(\Theta (t))}{\partial \theta _j}\right \rangle, \quad V_i= \text{Im} \left\langle \frac{\partial \psi(\Theta (t))}{\partial \theta _i} \right| \mc H \left | \psi(\Theta (t))\right \rangle
\end{equation}
can be measured on quantum circuits utilizing Hadamard test technique directly. The detailed measurement procedure and derivation are provided in  \cref{appE}. Thereafter, the EOM \cref{eq:eom} can be solved numerically with classical device. 
Moreover, \rev{since the circuit state remains normalized, we need to track the norm classically in order to recover the unnormalized state, especially along the evolution of linear QSD.}
% as in the case of linear QSD, the wavefunction is not always normalized, and additional tracking of its norm is required in VQS algorithm. %According to \citet{LuoLinGao2024}, 
The time derivative of the norm $\Gamma$ is given by
       \begin{equation}
           \dot{\Gamma}= \frac{\Gamma}{2\mathrm{i}}\langle \psi(\Theta(t)) |\mathcal{H}-\mathcal{H}^\dagger | \psi(\Theta(t))\rangle,
        \end{equation}
        which could be solved in the same manner as the EOM. 
        
In our case, we employ the fourth order Runge-Kutta (RK4) method, and update the parameters in the variational ansatz (along with the norm, if necessary) iteratively according to the above
equations. The parameterized quantum circuit obtained at each step will serve as the initial state for the next step, and the parameters will be updated sequentially. Notably, the original HVA may exhibit redundancy in its expressibility, especially for systems with a large number of Hamiltonian terms, where the HVA circuit can become prohibitively deep. Therefore, in some cases, we could employ a simplified HVA by removing some quantum gates without compromising the task.

\section{Lindblad simulation via stochastic Magnus expansion}\label{sec:stomagnus}

\subsection{Overview}\label{sec:overview}

The stochastic Magnus expansion lies at the heart of this study, providing a systematic way to solve non-commutative systems of SDEs, as in QSD unraveling of LME. To facilitate the discussion, we begin with a brief overview of the classical Magnus expansion in the deterministic version. For the linear ordinary differential equation of the form
\begin{equation}\label{eq:ode}
    \dv{Y(t)}{t}=A(t)Y(t), \quad Y(t_0)=Y_0
\end{equation}
with $A(t)$ a $n\times n$ coefficient matrix, the solution can be easily obtained as $Y(t)=\text{exp}(\int_0^t A(s)\dd s)Y_0$ when the matrix-valued continuous function $A(t)$ satisfies $[A(t_1),A(t_2)]=0$ for any values of  $t_1,t_2$. However, in most cases, this condition does not hold. It was Magnus \cite{Magnus1954} who first wrote down the general solution $Y (t) = \exp(\Omega(t)) Y_0$ for \cref{eq:ode} in non-commutative case as the ``\emph{continuous analogue}'' of the Baker-Campbell-Hausdorff (BCH) formula, where $\Omega(t)$ is expressed as an infinite series with its $p$-th partial sum denoted by $\Omega^{[p]}$, i.e.
\begin{equation}
    \Omega (t)=\lim_{p\to\infty} \Omega^{[p]}(t).
\end{equation}
The explicit expressions of first four partial sums are as follows~\cite{BlanesCasasOteo2009}:
\begin{equation}\label{eq:determagnus}
    \begin{aligned}
\Omega^{[1]}(t)&=  \int_0^t A\left(s_1\right) \mathrm{d} s_1,\\
\Omega^{[2]}(t)&= \Omega^{[1]}(t)+\frac{1}{2}\int_0^t\int_0^{s_1} [A(s_1), A(s_2)]\dd s_2 \dd s_1, \\
\Omega^{[3]}(t)&=\Omega^{[2]}(t)+\frac{1}{6}\int_{0}^{t} \int_0^{s_1} \int_0^{s_2}\left( \big[A(s_1),[A(s_2),A(s_3)]\big]+\big[[A(s_1),A(s_2)],A(s_3)\big]\right) \dd s_3 \dd s_2 \dd s_1 ,\\
\Omega^{[4]}(t)&=\Omega^{[3]}(t)+\frac{1}{12}\int_0^t\int_0^{s_1}\int_0^{s_2}\int_0^{s_3}\Bigg(\Big[\big[[A(s_1),A(s_2)],A(s_3)\big],A(s_4)\Big]+\Big[A(s_1),\big[[A(s_2),A(s_3)],A(s_4)\big]\Big]\\
&\quad +\Big[A(s_1),\big[A(s_2),[A(s_3),A(s_4)]\big]\Big]+\Big[A(s_2),\big[A(s_3),[A(s_4),A(s_1)]\big]\Big]\Bigg) \dd s_4 \dd s_3 \dd s_2 \dd s_1.
\end{aligned}
\end{equation}

Undoubtedly, there exist several alternatives to providing exact solutions formally of \cref{eq:ode}. For example, the Neumann expansion, commonly known in physics as the \emph{Dyson series}, offers an exponential-free framework for ODEs. However, for numerical simulations, the truncated Neumann series fails to preserve the Lie-group structure~\cite{Iserles2004}, which could be a considerable drawback. Moreover, other Lie-group-based expansions, such as \textit{Fer expansion}~\cite{BlanesCasasOteo2009}, are often complicated and challenging to implement, whereas the Magnus expansion lends itself to systematic improvements and can be directly generalized to stochastic scenarios.

In this context, we now turn our attention to the following general multi-dimensional Stratonovich-type SDE of the form:
\begin{equation}\label{eq:multidimsde}
    \dd \bv X_t  = G_0 \bv X_t \dd t +\sum_{j=1}^d   G_j\bv X_t\circ \dd W^j_t .
\end{equation}
Here, $\bv X_t\in \mathbb C^n$ is a $n$-dimensional random process, hollow circle $\circ$ denotes the Stratonovich interpretation of the SDE, $\{W^j_t\}_{j=1}^d$ are $d$ independent real-valued Wiener processes, and ~$G_0,  G_1\cdots, G_d\in \mathbb C^{n\times n}$ \cite{BurrageBurrage1999, Burrage1999,GriggsBurrageBurrage2024}. Analogous to the deterministic ODE case \cref{eq:ode}, the solution of the SDE \cref{eq:multidimsde} can also be formulated in the framework of an exponential integrator. Conceptually, this formulation involves replacing the deterministic integrals in \cref{eq:determagnus} with their stochastic counterparts. Specifically, we could formally define 
\begin{equation}\label{eq:formalAt}
    A(t)\dd t = G_0\dd t + \sum_{j=1}^d G_j\circ \dd W_t^j,
\end{equation}
and plug it back into the deterministic expression \cref{eq:determagnus} to derive the stochastic version. This approach, known as the \emph{stochastic Magnus expansion}, yields the corresponding Magnus operator $\Omega(t)$ as established in \cite{Burrage1999,BurrageBurrage1999}, which is given by
\begin{equation}\label{eq:stomagnus}
\begin{aligned}
\Omega(t)= & \sum_{j=0}^d G_j J_j+\frac{1}{2} \sum_{i=0}^d \sum_{j=i+1}^d[G_i, G_j](J_{j i}-J_{i j}) \\
& +\sum_{i=0}^d \sum_{k=0}^d \sum_{j=k+1}^d[G_i,[G_j, G_k]]\left(\frac{1}{3}(J_{k j i}-J_{j k i})+\frac{1}{12} J_i(J_{j k}-J_{k j})\right)+\cdots .
\end{aligned}
\end{equation}
Here $J_{j_1j_2\cdots j_k}$ denote the Stratonovich multiple integral, where the integration is taken with respect to $\dd t$ if $j_i=0$ or $\dd W_t^j$ if $j_i=j$. That is to say,
\begin{equation}
    J_{j_1j_2\dots j_k,t}\equiv \int_{0}^t\cdots \int_{0}^{s_3}\int_0^{s_2}\circ\, \dd W_{s_1}^{j_1}\circ \dd W_{s_2}^{j_2}\cdots \circ \dd W_{s_k}^{j_k},\quad \dd W_{t}^{0}\equiv \dd t.
\end{equation}
To put it concisely, for a multi-index $\alpha$, the Stratonovich integral $J_\alpha$ is just a certain Brownian random process. It is worth highlighting that $\{J_\alpha\}$ are inherently interdependent of one another, governed by a set of interrelations. Each can be systematically computed through the Fourier expansion technique of the Brownian bridge, as discussed in \cite{KloedenPlaten2013}. A comprehensive discussion of the stochastic Magnus expansion and multiple Stratonovich integrals is provided in \cref{appC}.

The convergence of Magnus expansion is crucial, and was studied by several previous works \cite{Moan2007, Casas2007, KevinStefanoAndrea2021}. In deterministic case, to the best of our knowledge, it has been proved that the sharpest bound of radius of convergence is given by 
\begin{equation}\label{eq:convergenceradius}
    r_c=\sup \left\{ t\geq 0\,\Big|\int_0^t \norm{A(s)} \, \dd s <\pi \right\}.
\end{equation} \citet{Moan2007} proved the case where $A(t)$ is a real matrix, while \citet{Casas2007} extended it to all bounded linear operator in Hilbert space. For the stochastic Magnus expansion, similar results could also be obtained, as established by \citet{KevinStefanoAndrea2021}. Therefore, in numerical simulations, we cannot take overly large step sizes, especially when the matrix has a large spectral radius. \rev{In practice, we typically choose the time step $\Delta $ of order $\mc O(r_c^{-1})$}.

\subsection{Magnus exponential integrators for solving QSD equations}

Since we need to derive high-order exponential integrator schemes for solving the QSD equations, it is helpful for us to transform the Itô-type SDEs in \cref{eq:nlqsd} and \cref{eq:linear} into the equivalent Stratonovich-type SDEs. This conversion can be achieved by applying results from Kunita \cite{Kunita1980, LiLi2019}. The resulting Stratonovich SDE takes the form\footnote{Strictly speaking, in our algorithmic implementation, the wavefunction trajectory is required to be normalized after each single step propagation via the sampled Magnus integrator according to \cref{eq:nlnqsd_stra}. We refer readers to \cref{appB} for detailed discussions.}
\begin{equation}\label{eq:nlnqsd_stra}   \dd\psi = -\ii {H}_s\psi \dd t+ \sum_k \left[2\Re (\langle L_k\rangle) L_k - \frac12 (L_k + L_k^\dagger )L_k \right]\psi \dd t + \sum_k {L}_k \psi \circ \mathrm{d}W_{t}^k \end{equation}
for nonlinear QSD, and 
\begin{equation}
    \dd \psi = -\ii H_s \psi \dd t -\frac12 \sum_k (L_k+L_k^\dagger)L_k \psi \dd t + \sum_k L_k \psi \circ \dd W_t
\end{equation}
for linear QSD. A comprehensive derivation of these transformations is provided in \cref{appB}.

Based on the truncated stochastic Magnus expansion, one can establish our explicit single step numerical methods. Recall that the Stratonovich integrals $J_i, J_{ij}, J_{ijk},\cdots$ are Brownian random processes. To apply the numerical schemes discussed in \cref{sec:overview}, we denote
\begin{equation}\label{eq:g0}
        \begin{aligned}G_{0,\text{linear}} = -\ii H_s - \frac12 \sum_k (L_k+L_k^\dagger)L_k , \quad G_{0,\text{nonlinear}} = -\ii {H}_s+ \sum_k \left[2\Re (\langle L_k\rangle) L_k - \frac12 (L_k + L_k^\dagger )L_k \right],
        \end{aligned}
\end{equation}
\begin{equation}
    G_{k,\text{linear}} = G_{k,\text{nonlinear}} = L_k\quad (k\ge 1).
\end{equation}
Hence at each time step with step length $\Delta$, we can first sample $J_\alpha$, $\Delta$ according to \cref{appC,appD}. Thanks to the independent increment property of the underlying Wiener process, this sampling procedure does not require retaining the full history of previous increments. The first-order truncation becomes
%Hence at each time step with step length $\Delta$, we can first sample $J_{\alpha,\Delta}$ according to \cref{appC,appD}, then the first-order truncation becomes
\begin{equation}\label{eq:omega1}
    \Omega^{[1]}(\Delta)=\sum_{j=0}^d G_jJ_j,
\end{equation}
and we denote this format as Scheme I. It is worth mentioning that even the Scheme I is different from the Euler-Maruyama method for QSD, due to the conversion between Itô- and Stratonovich-type SDE. Similarly, we construct the Scheme II, III and IV as \footnote{Note that the interrelationships between multiple stochastic integrals, as given in \citet{BurrageBurrage1999, Burrage1999}, can also be used to reduce the number of stochastic integrals required for evaluating $\Omega^{[4]}$ in Scheme IV, similar to Scheme III. However, its expression is highly intricate, so we do not provide its explicit form. An illustrative derivation of $\Omega^{[4]}$ in RPM is available in \cref{appC,appD}.

}
\begin{equation}\label{eq:omega2}
    \Omega^{[2]}(\Delta)=\Omega^{[1]}(\Delta)+\frac{1}{2} \sum_{i=0}^d \sum_{j=i+1}^d[G_i, G_j](J_{j i}-J_{i j}),
\end{equation}
\begin{equation}\label{eq:omega3}
    \Omega^{[3]}(\Delta)=\Omega^{[2]}(\Delta)+\sum_{i=0}^d \sum_{k=0}^d \sum_{j=k+1}^d[G_i,[G_j, G_k]]\left(\frac{1}{3}(J_{k j i}-J_{j k i})+\frac{1}{12} J_i(J_{j k}-J_{k j})\right),
\end{equation}
\begin{equation}\label{eq:omega4main}
    \Omega^{[4]}(\Delta) = \Omega^{[3]}(\Delta) +\sum_{i,j,k,l=0}^d  \Big[\big[[G_i,G_j],G_k\big],G_l\Big]\left(\frac{1}{12}(J_{lkji}-J_{kjil}+J_{jikl}+J_{iklj})\right).
\end{equation}

We would like to point out that, although the higher-order truncation seems to include a larger number of additional commutators of $H$ and $L_i$, the sampling cost of the effective Hamiltonian does not grow without bound. During algorithm implementation, the effective Hamiltonian is classically decomposed into non-redundant Pauli strings. Consequently, despite the apparent complexity of higher-order schemes, the measurement cost remains upper bounded by a system-size-dependent quantity. Specifically, in a system of size $N$, where $\lceil\log_2N\rceil$ qubits are used for simulation, \rev{the number of Pauli strings that need to be measured is at most $
    4^{\lceil\log_2N\rceil}\le  N^2.$}
Although this upper bound may initially appear comparable to the complexity of superoperator-based propagation as illustrated in \cref{fig:illus}(a), the actual computational cost is fundamentally different. In contrast to the superoperator approach—which may involve substantial hidden overhead—the Pauli measurements can be executed efficiently in parallel, greatly reducing the practical burden of implementation. For systems with specific forms of jump operators and Hamiltonians, the number of additional measurement terms introduced by higher-order commutators may be quite limited, as observed in the radical pair model. Furthermore, the higher-order nested commutators always appear alongside higher powers of $\Delta$. As a result, when $\Delta$ lies within the convergence radius of the expansion, some of the additional Pauli strings may carry coefficients with negligible contributions, allowing the sampling cost to be further reduced by discarding them.
\subsection{Runge-Kutta-Munthe-Kaas (RKMK)-type nonlinear correction}\label{sec:corr}

In one-step methods, the numerical solution at the next step depends only on the current value. In the nonlinear QSD, the standard approach to deal with the nonlinear term $\langle L_k \rangle = \psi_t^\dagger L_k \psi_t$ is to apply the approximation $\langle L_k \rangle_{t\in [t_0, t_0+\Delta]}\approx \langle L_k \rangle_{t=t_0} $. That is, we first compute $G_{0,\text{nonlinear}}$ in \cref{eq:g0} using $\psi_0$ as
\begin{equation}
    G_{0,\text{nonlinear}} = G_{0,\text{nonlinear}}(\psi_0) =  -\ii {H}_s+ \sum_k \left[2\Re (\psi_0^\dagger  L_k \psi_0) L_k - \frac12 (L_k + L_k^\dagger )L_k \right]
\end{equation}
and then use this $G_{0,\text{nonlinear}}(\psi_0)$ to derive the Magnus integrators, denoted as $\Omega(\Delta;\psi_0)$, according to \cref{eq:omega1,eq:omega2,eq:omega3,eq:omega4main}. 
However, this approximation may lead to substantial cumulative errors, particularly when the dynamics undergoes significant variations within the chosen step size interval $\Delta$. To mitigate this potential drawback, we propose an alternative scheme for nonlinear QSD employing a nonlinear correction with a form that resembles the Runge-Kutta-Munthe-Kaas (RKMK) Heun method. This correction also appears quite straightforwardly in the explicit Magnus expansions for nonlinear equations demonstrated in \citet{CasasIserles2006}. Next we will outline the theory behind this type of nonlinear correction.
%To derive this, we consider a single evolution step using the Magnus integrator $\Omega(t,\psi_0)$. 

Without loss of generality, we assume an initial state $\psi_0$ and a time interval $[0,\Delta]$. We formally define the non-Hermitian Hamiltonian as
\begin{equation}
    A(t,\psi_t) \dd t  = -\ii \mc H(t,\psi_t) \dd t.
\end{equation}  
%In Magnus integrators of any order, 
Note that the evolution involves only Wiener random processes. Consequently, for a given wavefunction trajectory, %\(\Omega(t)\) 
$\mc H(t, \psi_t)$ can be sampled and realized, and the sampled path is almost surely continuous. Therefore, upon sampling, we treat the non-Hermitian Hamiltonian \(\mathcal{H}\) as a deterministic operator dependent on \(t\) and \(\psi_t\) for \(t\in [0,\Delta]\), while the wavefunction \(\psi_t\) can be regarded as a deterministic vector expressed in an integral form. Specifically, up to \(\mathcal{O}(\Delta^2)\), we obtain
\begin{equation}
    \psi_t = \psi_0  -\ii \int_0^t \mathcal{H}(s, \psi_s) \psi_0 \, \dd s.
\end{equation}  
Applying the two-point trapezoidal rule for discretization, we obtain
\begin{equation}
%\psi_\Delta =\psi_0 -\frac{\ii \Delta }{2}\left(\mathcal{H}(0,\psi_0)+\mathcal{H}(\Delta, e^{\Omega(\Delta,\psi_\Delta )}\psi_0)\right)\psi_0.
\psi_\Delta =\psi_0 -\frac{\ii \Delta }{2}\left(\mathcal{H}(0,\psi_0)+\mathcal{H}(\Delta,\psi_\Delta )\right)\psi_0.
\end{equation}

At first glance, determining the explicit form of %$\Omega(\Delta,\psi_\Delta)$ 
$\psi_\Delta$ appears necessary. However, to construct a numerical scheme with accuracy $\mathcal{O}(\Delta^2)$, it suffices to employ a first-order approximation of it. For instance, by applying the Euler method, we derive an alternative explicit scheme of order $\mathcal{O}(\Delta^2)$ that depends only on $\psi_0$, %\HL{It could be further justified}
\begin{equation}
\psi_\Delta = \psi_0 -\frac{\ii \Delta }{2}\left(\mathcal{H}(0,\psi_0)+\mathcal{H}(\Delta, e^{\Omega(\Delta;\psi_0)}\psi_0)\right)\psi_0 + \mathcal{O}(\Delta^2) = \exp (\widetilde{\Omega}(\Delta))\psi_0 +\mathcal{O} (\Delta^2).
\end{equation}
Here,
\begin{equation}
\widetilde{\Omega} (\Delta)= \frac{1}{2}\left( \Omega(\Delta ; \psi_0) + \Omega (\Delta; e^{\Omega(\Delta;\psi_0) }\psi_0)\right).
\end{equation}
%This formulation enhances theoretical clarity while maintaining the desired accuracy, eliminating the need for an explicit evaluation of $\Omega(\Delta,\psi_\Delta)$.
%to better preserve the structure of the nonlinear dynamics \cite{CasasIserles2006}. It can be expressed as
%\begin{equation}
%    \Omega^\prime (t)= \frac{\Omega(0,\mathbf{X}_0)+\Omega(\Delta, e^{\Omega(0,\mathbf{X}_0)}\mathbf{X}_0)}{2} ,
%
%\end{equation}
Then, we can update $\psi_0$ to $\psi_\Delta$ using the corrected exponential integrator $\widetilde \Omega(\Delta )$
\begin{equation}
    \psi_{\Delta} = e^{\widetilde \Omega(\Delta )} \psi_0.
\end{equation}
It is important to note that higher-order schemes can yield more accurate results, but they often involve increased computational complexity, such as additional exponential calculations (in classical algorithms) or more variational steps (in quantum algorithms). In some cases, this may be less efficient than simply reducing the length of the time step and a trade-off must be considered. Nonetheless, in most cases, the one or two-step method is already practically adequate. The improvement provided by the RKMK-type nonlinear correction is numerically verified in \cref{sec:numerics}.

\section{Numerical results}\label{sec:numerics}

In this section, we employ the proposed algorithm to precisely simulate the dynamics of three distinct systems. For the transverse field Ising model (TFIM) with damping, we first demonstrate that higher-order Magnus methods significantly improve accuracy for larger time steps and extended simulation durations. Leveraging nonlinear QSD, both the TFIM and the Fenna-Matthews-Olson (FMO) complex exhibit exceptional numerical stability and substantial variance reduction compared to linear schemes. Furthermore, through the precise simulation of the FMO complex, we illustrate the strong corrective effect of the multi-step method on the nonlinear characteristics of the equations. Finally, we conduct a preliminary validation of our approach on the radical pair model (RPM) model, where the higher-order Magnus methods also demonstrate their advantages. The corresponding variational circuit implementations are detailed in \cref{appE}, and parameter settings used for our numerical experiments are summarized in \cref{appF}. Our codes for the numerical experiments are available via GitHub \cite{github:LindbladMagnus}.

Unless otherwise stated, the \emph{exact calculations} of Lindblad equations in the following numerical experiments are performed using \texttt{QuTiP}. In each simulation, we sample and propagate the wavefunction trajectory ensemble $ \{\psi_k(t)\}_{k=1}^{N_{\text{traj}}} $ with a total of $ N_{\text{traj}} $ trajectories and compute the ensemble-averaged expectation values of observables of our interest, such as the overlap with specific quantum states. Obviously, this is equivalent to use the ensemble average of the wavefunctions to approximate the many-body density operator under the evolution of Lindblad equations
\begin{equation}
    \rho (t)  \approx \frac{1}{N_{\text{traj}}}\sum_{\ell =1}^{N_{\text{traj}}} \psi_\ell (t) \psi_\ell (t) ^\dagger
\end{equation}
and evaluate the expectation value $\Tr(\rho O)$ for some certain observable $O$. 
\subsection{Transverse field Ising model (TFIM) with damping}\label{sec:tfim}

We start by examining the performance of our algorithm on dissipative dynamics of TFIM with two sites. The general Hamiltonian of TFIM is given by
\begin{equation}
    {H}_s=J\sum_{i}\sigma^i_z\sigma^{i+1}_{z}-h\sum_{i} \sigma_x^i,
\end{equation}
where $J$ is the coupling strength and $h$ is the transverse magnetic field. The jump operator $L_k$ for site $k$ is set to $L_k=\sqrt{\Gamma_k}(\sigma_x^k+\ii \sigma_y^k)/2$. The corresponding HVA can be expressed as
\begin{equation}
    U_{\text{TFIM}}(\Theta(t))=\prod_{p=1}^m e^{- \frac{\ii}{2}\theta_{7,p}\sigma_z^1\sigma_z^2}e^{- \frac{\ii}{2}\theta_{6,p}\sigma_z^1}e^{- \frac{\ii}{2}\theta_{5,p}\sigma_z^2}e^{- \frac{\ii}{2}\theta_{4,p}\sigma_y^1}e^{- \frac{\ii}{2}\theta_{3,p}\sigma_y^2}e^{- \frac{\ii}{2}\theta_{2,p}\sigma_x^1}e^{- \frac{\ii}{2}\theta_{1,p}\sigma_x^2}.
\end{equation}

\begin{figure}[htbp]
    \centering
    \includegraphics[width=0.9\linewidth]{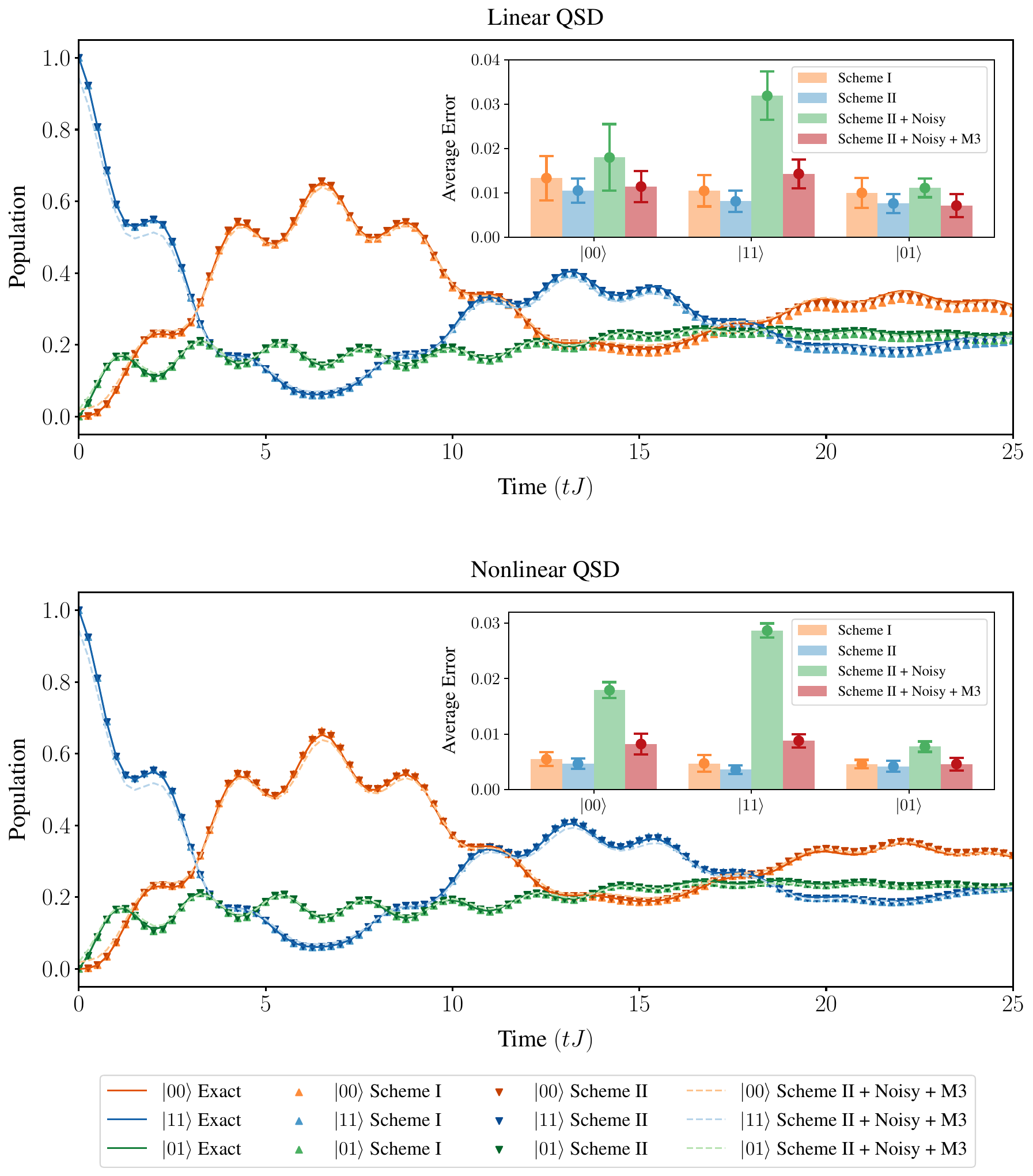}
    \caption{\raggedright Performances of the numerical methods based on linear QSD (the upper panel) versus nonlinear QSD (the lower panel) and also Scheme I versus Scheme II Magnus integrators. Noisy test results obtained from quantum circuit executions on the \texttt{FakeBelemV2} fake backend, both with and without error mitigation using the M3 package, are also shown. All results from the IBM fake backend are obtained with $2^{19}$ shots. We plot the evolution of each state population $\Tr(\rho \dyad{00})$, $\Tr(\rho\dyad{11})$ and $\Tr(\rho\dyad{01})$, where $\rho \approx \frac{1}{N_{\text{traj}}}\sum_{\ell =1}^{N_{\text{traj}}} \psi_\ell\psi_\ell^\dagger $. We set the step length $\Delta=0.25$ $tJ$, trajectory number $N_{\text{traj}} = 10^3$ and the stopping time $T = 25~tJ$. We repeat the experiment for $10$ times and evaluate the average error of the populations at each step throughout the simulation. The inset shows the average error for each scheme and each state, with error bars representing the 99\% confidence interval.}
    \label{fig:TFIM_linear_nonlinear_comp}
\end{figure}

In our simulation, we set $J=1, \Gamma_k = 0.1$ for all $k$, and the number of layers in HVA is $m=3$. The initial system is chosen as the pure state $|11\rangle$ where both spins are in the spin-up state. We propagate the system dynamics using a time step of $\Delta =0.25~tJ$, which is considerably larger compared with \citet{LuoLinGao2024}. We adopt the number of trajectories $N_{\text{traj}} = 10^3$ and simulate the system dynamics up to the stopping time of $T= 25~tJ$. In order to reduce statistical bias and better demonstrate robustness, we repeat the experiment 10 times with different random seeds. 
 We track the population evolution of the $ |00\rangle $, $ |01\rangle $, and $ |11\rangle $ states, with the results obtained using Scheme I and Scheme II for both linear (upper panel) and nonlinear QSD (lower panel) presented in \cref{fig:TFIM_linear_nonlinear_comp}. Additionally, we evaluate the average error in the overlap with the states at each step throughout the dynamics, with the results displayed in the insets of the respective panels. Our observations indicate that, in both the linear and nonlinear cases, Scheme II achieves better agreement with the exact solution than Scheme I. Moreover, the nonlinear unraveling exhibits superior accuracy compared to the linear approach, validating the advantage of the norm-preserving property of nonlinear QSD discussed in \cref{sec:unravel} and \cref{appA}.

To further validate the applicability of our algorithm on near-term quantum devices, we perform additional simulations under settings of realistic noisy quantum hardware. An instructive observation is that, owing to the inherently stochastic nature of our unraveling-based approach, the result is obtained by averaging over multiple wavefunction trajectories evolved according to stochastic differential equations. Consequently, the impact of noise is effectively averaged out, which imparts the method with an intrinsic robustness to noise. 

In our experiments, we adopt a setup similar to that in \citet{ChenGomesNiuEtAl2024}. In their setup, the variational parameters of the time evolution are computed classically, after which the quantum circuits are executed on a real quantum computer to measure observables. Following this methodology, we utilize the fake backend provided by the \texttt{Qiskit-IBM-Runtime} package, which mimics the behavior of IBM Quantum systems using system snapshots. These system snapshots record the essential information about the quantum system and can be used to perform noise simulations. Our simulations are run on \texttt{FakeBelemV2}, a fake $5$ qubit backend with a \emph{T-shaped} coupling topology. We also apply readout error mitigation using the matrix-free measurement mitigation (M3) package, which provides effective correction of noisy measurement samples. For each measurement task, we perform $2^{19}$ shots. We also repeat the experiment 10 times, each consisting of $10^3$ trajectories, as in the noiseless scenario. The population evolution trajectories and average errors are also shown in \cref{fig:TFIM_linear_nonlinear_comp}. The simulation results show great agreement with the exact solution, indicating that the QSD method is relatively noise-tolerant. Moreover, the applied readout error mitigation scheme proves to be effective in correcting measurement errors.

%\begin{figure}[htbp]
%    \centering
%    \includegraphics[width=0.9\linewidth]{figure/TFIM_noisy_ideal_comp.pdf}
%    \caption{\raggedright Performances of the numerical methods based on linear QSD (the upper panel) versus nonlinear QSD (the lower panel) and also Scheme I versus Scheme II Magnus integrators. We plot the evolution of each state population $\Tr(\rho \dyad{00})$, $\Tr(\rho\dyad{11})$ and $\Tr(\rho\dyad{01})$, where $\rho \approx \frac{1}{N_{\text{traj}}}\sum_{\ell =1}^{N_{\text{traj}}} \psi_\ell\psi_\ell^\dagger $. We set the step length $\Delta=0.25$ $tJ$, trajectory number $N_{\text{traj}} = 10^3$ and the stopping time $T = 25~tJ$. We repeat the experiment for $10$ times and evaluate the average error of the populations at each step throughout the simulation. The inset shows the average error for each scheme and each state, with error bars representing the 99\% confidence interval.}
%    \label{fig:TFIM_linear_nonlinear_comp}
%\end{figure}
\subsection{Fenna-Matthews-Olson (FMO) complex}\label{sec:fmo_numer}

\begin{figure}[htbp]
    \centering
    \includegraphics[width=0.5\linewidth]{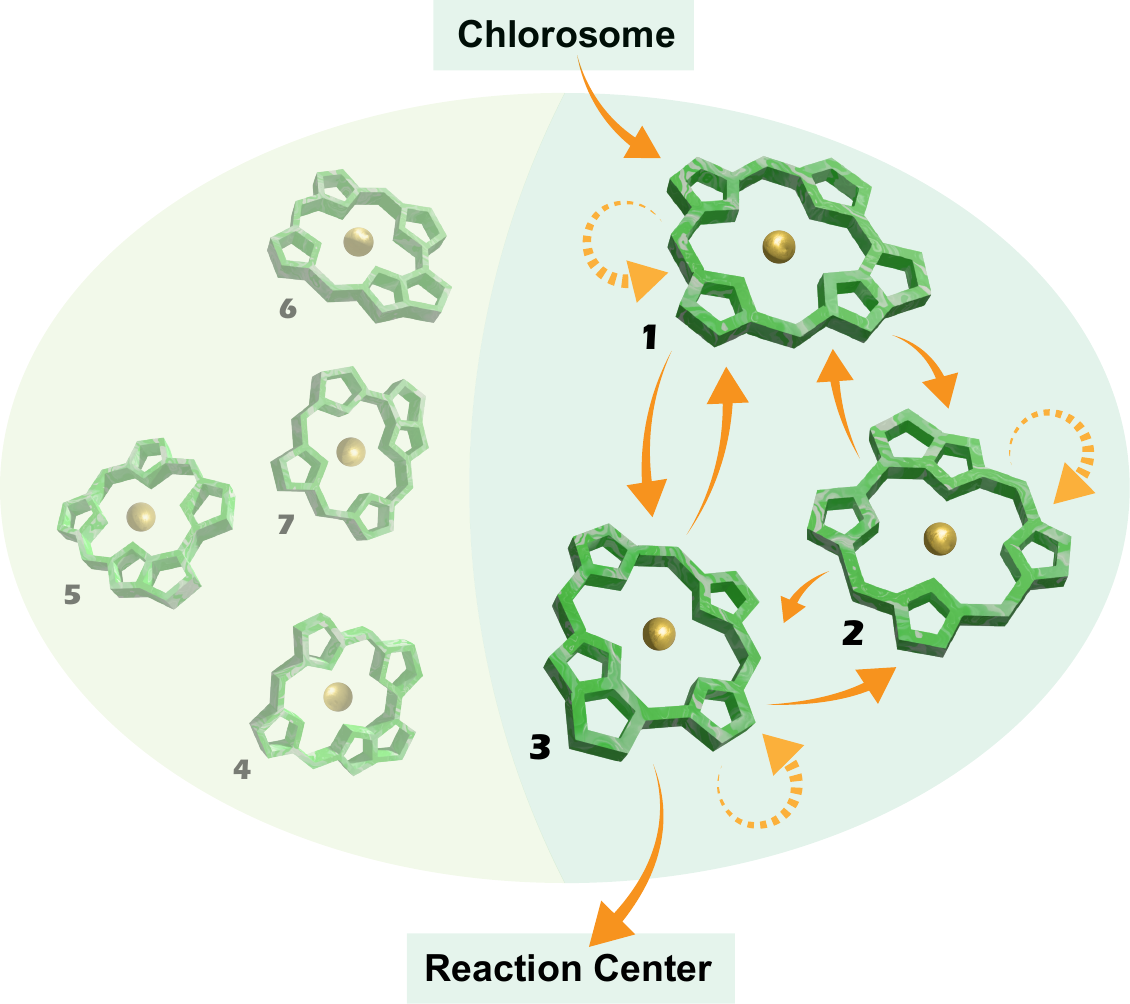}
    \caption{\raggedright Conceptual illustration of the FMO model. In the FMO complex, the chromophores 1 to 7 facilitate the transfer of excitation energy from the chlorosome antennae to the reaction center. We consider the simplified 3-site system and model this quantum dissipative dynamics using Lindblad equation, with Hamiltonian defined in \cref{eq:Ham_FMO,eq:Ham_matrix_FMO} and jump operators defined in \cref{eq:FMO_deph,eq:FMO_diss,eq:FMO_sink}.}
    \label{fig:FMO_concept}
\end{figure}

The Fenna-Matthews-Olson (FMO) complex is a well-studied pigment–protein complex (PPC) found in green sulfur bacteria, playing a crucial role in photosynthetic light harvesting. It functions as a quantum wire, facilitating the transfer of excitation energy from large chlorosome antennae to the membrane-embedded bacterial reaction center, where charge transfer is initiated. Structurally, FMO exists as a trimer complex, with each monomer containing seven bacteriochlorophyll chromophores. As conceptually illustrated in \cref{fig:FMO_concept}, excitation typically occurs at chromophore 1 or 6 and is transported to chromophore 3, which is tightly connected with the reaction center. This energy transfer is influenced by environmental interactions, including neighboring excitations and the protein scaffold, and follows multiple quantum pathways, ensuring efficient exciton transport. While extensively studied both theoretically and experimentally, quantum algorithm-based investigations remain limited, and simulating the full 7-site system along with long-time dynamics remains a challenge.

\begin{figure}[htbp]
    \centering
    \includegraphics[width=0.80\linewidth]{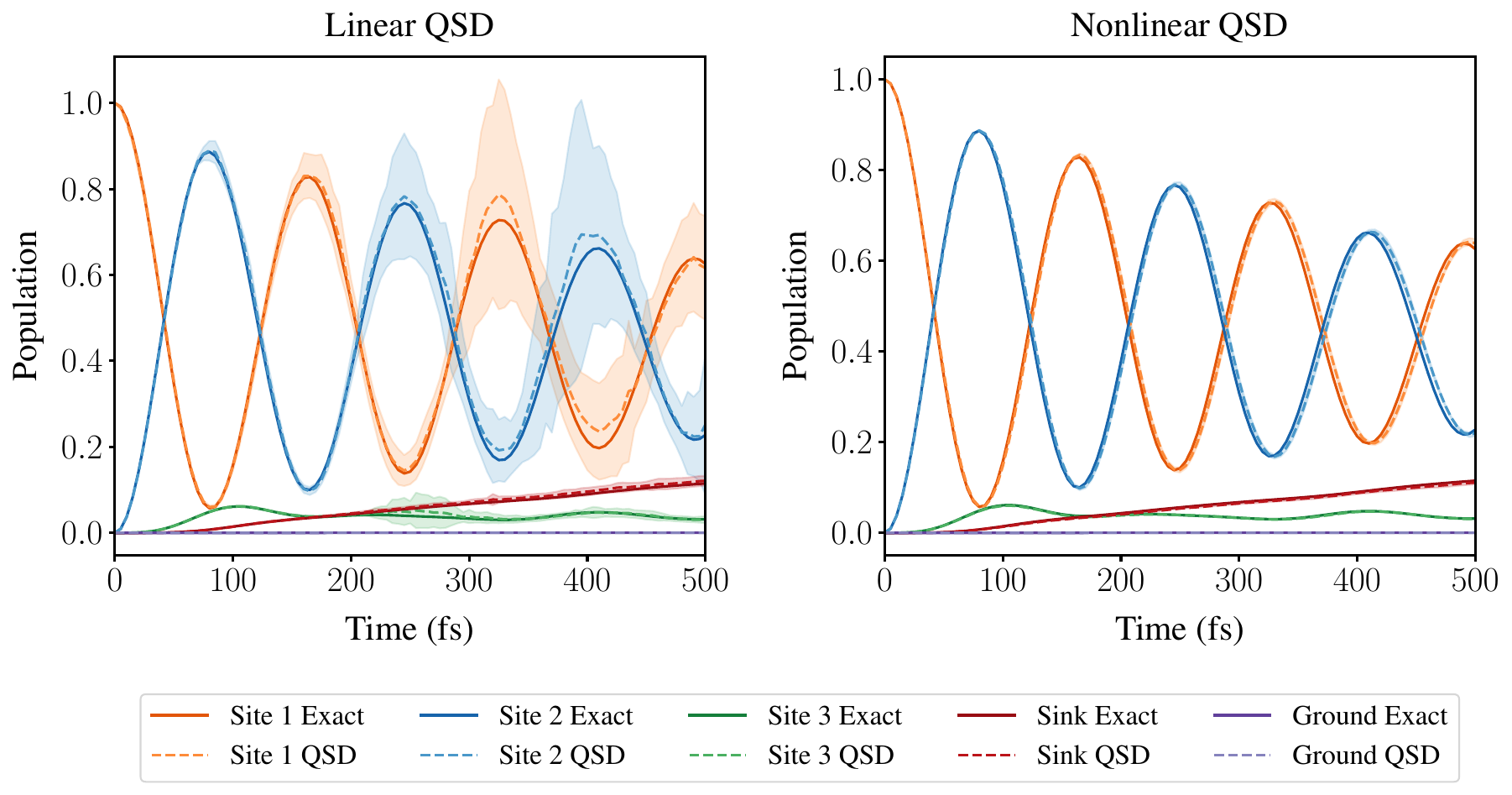}
    \caption{\raggedright
   Examining the performance of the numerical schemes based on linearly and nonlinearly unraveled Lindblad dynamics. We use the Scheme I for both tests and track the evolution of the expectation values $\Tr(\rho \dyad i)$ for $\rho \approx \frac{1}{N_{\text{traj}}}\sum_{\ell =1}^{N_{\text{traj}}} \psi_\ell\psi_\ell^\dagger$ of the states $\dyad1$ (site 1), $\dyad 2$ (site 2), $\dyad 3$ (site 3), $\dyad 4$ (sink) and $\dyad 0$ (ground). In each simulation, we use $N_{\text{traj}} = 10^3$, $\Delta = 5$ fs and the stopping time $T = 500$ fs. We repeat the experiment $10$ times with different random seeds, and the shaded area represents the 99\% confidence interval obtained from this sample.}
    \label{fig:FMO_linear_nonlinear}
\end{figure}

\begin{figure}[htbp]
    \centering
    \includegraphics[width=0.53\linewidth]{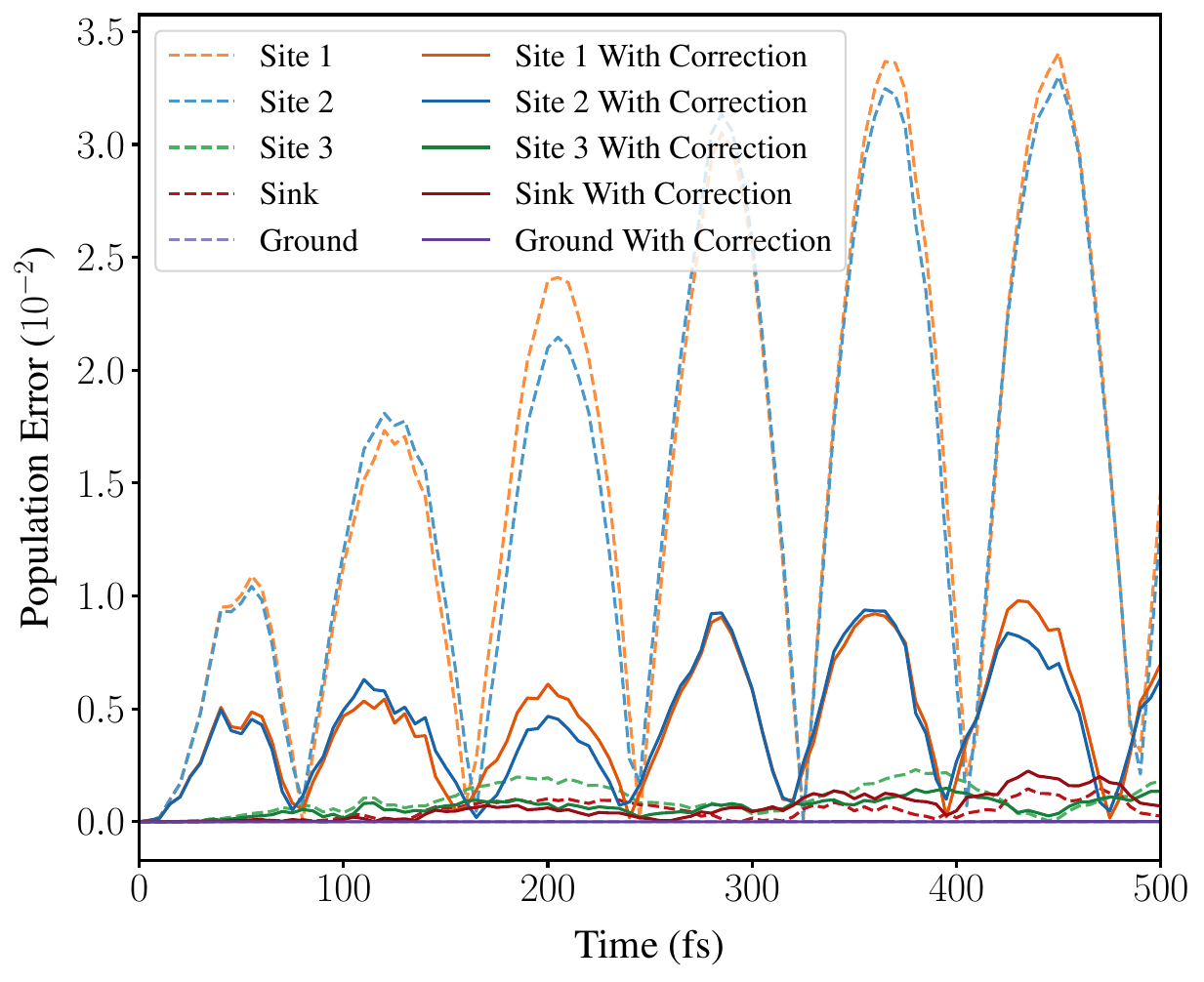}
    \caption{\raggedright
    The test of the performance of the RKMK-type nonlinear correction demonstrated in \cref{sec:corr}. We use $N_{\text{traj}} = 10^4$ for each test. We can observe that the error in the population of each interested state is significantly reduced, and the rapid growth of error with step number is mitigated effectively.
 }
    \label{fig:FMO_nonlinear_correction}
\end{figure}
The Hamiltonian of FMO complex is 
\begin{equation}\label{eq:Ham_FMO}
    H_s = \sum_{i=0}^4 \omega_i \sigma_i^+ \sigma_i^- +\sum_{j\ne i} J_{ij}(\sigma_i^+\sigma_j^-+\sigma_j^+\sigma_i^-).
\end{equation}
Here $\omega_i$ is the energy of the state $\ket{i}$, $J_{ij}$ is the coupling strength of the states $\ket i $ and $\ket j$, and $\sigma_i^+$, $\sigma_i^-$ denote the Pauli raising operator and Pauli lowering operator with respect to the state $\ket i$, respectively. In this paper, we adopt the parameters in \citet{HuHeadMarsdenMazziottiEtAl2022} (see \cref{appF} for details), which gives a Hamiltonian with the following explicit matrix form in the unit of eV
\begin{equation}\label{eq:Ham_matrix_FMO}
    H_s = \left(\begin{array}{ccccc}
0 & 0 & 0 & 0 & 0 \\
0 & 0.0267 & -0.0129 & 0.000632 & 0 \\
0 & -0.0129 & 0.0273 & 0.00404 & 0 \\
0 & 0.000632 & 0.00404 & 0 & 0 \\
0 & 0 & 0 & 0 & 0
\end{array}\right).
\end{equation}

For the set of jump operators $\{L_i\}_{i=1}^7$ in the FMO complex model, $L_1$ through $L_3$ are the dephasing operators 
\begin{equation}\label{eq:FMO_deph}
    L_{\text{deph},i } = \sqrt{\alpha} \dyad{i},\quad i = 1,2,3.
\end{equation}
$L_4$ through $L_6$ are dissipation operators then describe the transition from $\ket{i}$ to the ground:
\begin{equation}\label{eq:FMO_diss}
    L_{\text{diss}, i} = \sqrt{\beta}\dyad{0}{i}, \quad i=1,2,3.
\end{equation}
Finally, $L_7 = L_{\text{sink}}$ is the sink operator that describe the transition from the state $\ket{3}$ to the sink $\ket 4$, i.e.
\begin{equation}\label{eq:FMO_sink}
    L_{\text{sink}} = \sqrt\gamma \dyad{4}{3}.
\end{equation}

%\cite{HuHeadMarsdenMazziottiEtAl2022} 

We initialize the system in the state $ \rho_0 = \dyad{1} $ and propagate the wavefunction ensemble using a time step of $ \Delta = 5 $ fs, up to a stopping time of $ T = 500 $ fs. In each test, we sample and simulate $ N_{\text{traj}} = 10^3 $ wavefunction trajectories and repeat the experiment 10 times with different random seeds to ensure statistical robustness. Scheme I is employed for both tests, and we track the evolution of the expectation values $ \Tr(\rho \dyad{i}) $, where the density matrix is approximated as $ \rho \approx \frac{1}{N_{\text{traj}}} \sum_{\ell =1}^{N_{\text{traj}}} \psi_\ell\psi_\ell^\dagger $, for $ i = 0,1,2,3,4 $.

By comparing the graphs on the left and right of \cref{fig:FMO_linear_nonlinear}, we observe that the nonlinear QSD-based Magnus integrator yields significantly more accurate results for each state population than the linear approach. This confirms the robustness ensured by the norm-preserving property discussed in \cref{sec:unravel} and \cref{appA}, while also demonstrating the effectiveness of the stochastic Magnus expansion-based numerical scheme. We can also observe that the Scheme I is already sufficient to provide a good agreement with the exact solution in the nonlinear case. 

In fact, the nonlinear QSD results can be further enhanced by applying an RKMK-type correction to the single-step method for solving nonlinear dynamics, as introduced in \cref{sec:corr}. As shown in \cref{fig:FMO_nonlinear_correction}, incorporating this correction reduces the error in each state population. Additionally, we observe that in the absence of correction, the error grows significantly with the number of steps. However, with the RKMK-type nonlinear correction, this issue is effectively mitigated, further improving the stability and accuracy of the method.

\subsection{Radical pair model (RPM) for avian compass dynamics}\label{sec:rpm_numer}

\begin{figure}[htbp]
    \centering
    \includegraphics[width=0.6\linewidth]{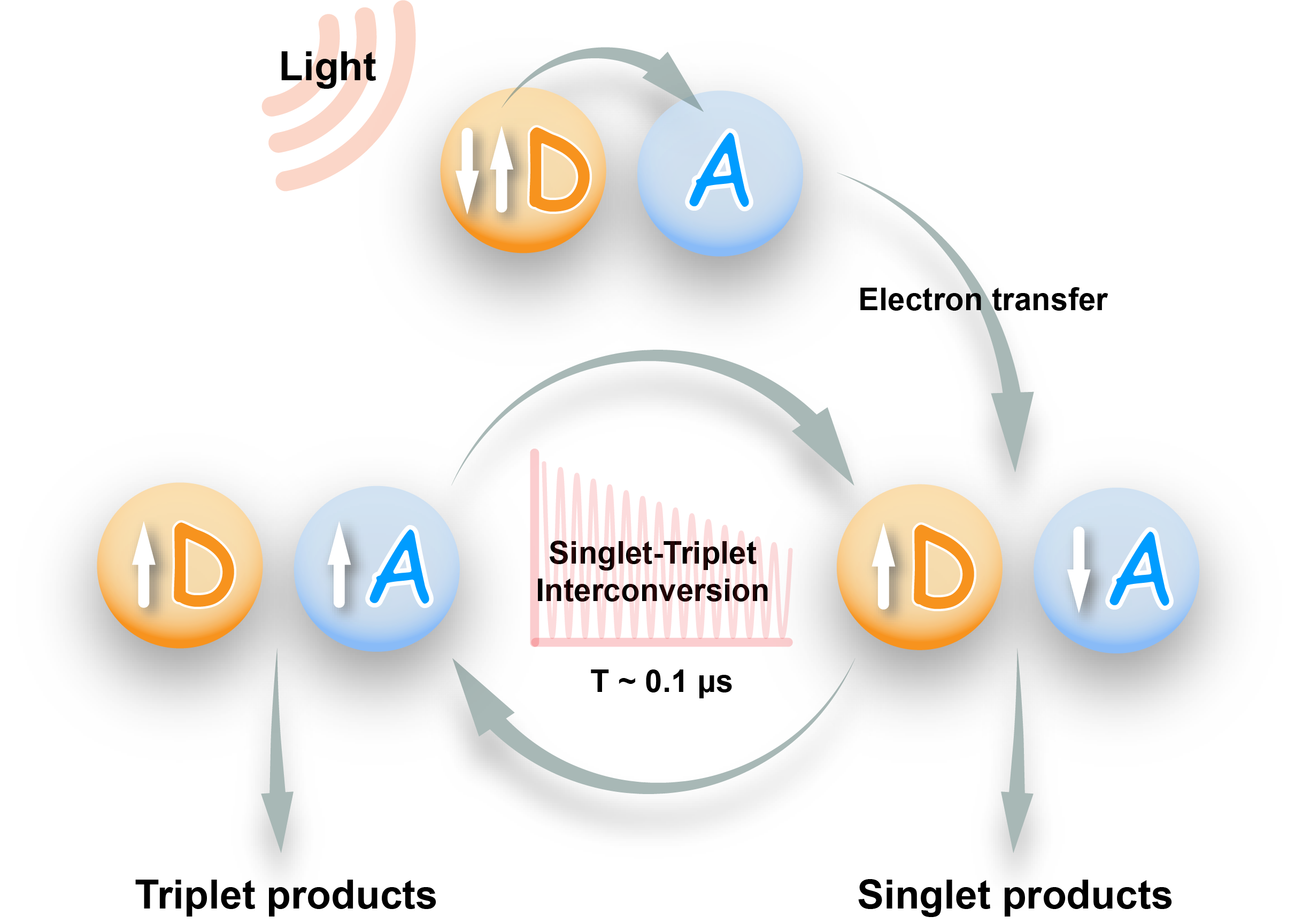}
    \caption{\raggedright
    Conceptual illustration of the radical pair model. The fast interconversion between the singlet and triplet radical pairs driven by the external geomagnetic field together with the internal hyperfine effect exhibits a highly oscillatory behavior with $T\sim 0.1~\text{\textmu}$s.}
    \label{fig:RPM_concept}
\end{figure}

As a non-trivial quantum effect in biology, avian compass exemplifies a remarkable natural phenomenon in which spin-selective radical pair dynamics enable highly sensitive magnetic perception, facilitating birds' ability to detect the geomagnetic field for orientation and navigation \cite{PaulsZhangBermanKais2013,Kominis2012}. In a nutshell, the radical pair mechanism comprises the following sequence: a photon initially excites a photosensitive molecule in the bird’s eyes, initiating an electron transfer reaction that generates a singlet radical pair. This radical pair subsequently undergoes interconversion between singlet and triplet states, governed by the interplay of the external geomagnetic field and intrinsic hyperfine interactions. Radical pairs in distinct spin configurations give rise to chemically distinguishable signals, enabling birds to infer directional information based on the geomagnetic field. This process is conceptually illustrated in \cref{fig:RPM_concept}. To analyze the quantum effects underlying radical pair reactions, we can model their dynamics using a Lindblad equation incorporating two additional \emph{shelving states} to quantify singlet and triplet yields, as demonstrated in \citet{GaugerRieperMortonBenjaminEtAl2011}.

Following \citet{PooniaKondabagilSahaGanguly2017}, we provide a concise introduction to the RPM model. The system Hamiltonian of RPM is 
\begin{equation}
    {H}_s = \frac{1}{\hbar}\left[\mu_Bg\mathbf{B}\cdot({\bv S}_1+{\bv S}_2)+{\bv I}\cdot \mathbf{A}\cdot {\bv S}_2\right],
\end{equation}
where the first term represents the coupling between the electron spin and the geomagnetic field, while the second term accounts for the coupling between the nuclear spin and the electron spin. Here $\bv S_1$ and $\bv S_2$ are Pauli spin operators $\bv S_i=(\sigma_x^i, \sigma_y^i, \sigma_z^i)$, $\bv I$ is the nuclear spin operator, and $\mathbf{A}=\text{diag}(a_x,a_y,a_z)$ denotes the hyperfine tensor. The geomagnetic field is characterized by $\mathbf{B}=B_0(\sin{\theta}\cos{\phi}, \sin{\theta}\sin{\phi},\cos{\theta})$. $\hbar$, $\mu_B$ and $g$ denotes the reduced Planck constant, Bohr magneton and the electron-spin $g$-factor. The angle $\theta$ denotes the orientation of magnetic
field. The jump operators are defined as follows:
\begin{equation}
\begin{aligned}
L_1&=\left|S\rangle\langle s, \uparrow\right|,& L_2&=\left|S\rangle\langle s, \downarrow\right|, \\ L_3&=\left|T\right\rangle\left\langle t_0, \uparrow\right|, &L_4&=\left|T\right\rangle\left\langle t_0, \downarrow\right|,\\
L_5&=\left|T\right\rangle\left\langle t_{+}, \uparrow\right|, &L_6&=\left|T\right\rangle\left\langle t_{+}, \downarrow\right|, \\
L_7&=\left|T\right\rangle\left\langle t_{-}, \uparrow\right| ,&L_8&=\left|T\right\rangle\left\langle t_{-}, \downarrow\right|.
\end{aligned}
\end{equation}

It is worth noting that the effective Hamiltonian $H_{\rm eff} = H_s -\frac\ii 2\sum_k L_k^\dagger L_k$ has a real part with a norm of $10^7$ order of magnitude. Intuitively, this suggests that the system is highly oscillatory. It resembles a harmonic oscillator with a very large ``undamped frequency'', resulting an oscillation period of approximately $T\sim 10^{-7}$ s. Consequently, due to the issue imposed by the radius of convergence in \cref{eq:convergenceradius}, it is necessary for us to adopt a relatively small time step, specifically $\Delta = 1\times 10^{-7}$s. Moreover, the pronounced oscillatory nature of this system implies that the expectation values $\langle L_k\rangle$ of the jump operators can  change rapidly even within a single time step. As a result, the nonlinear unraveling approach is not applicable for achieving an accurate simulation.

\begin{figure}[htbp]
    \centering
    \includegraphics[width=0.80\linewidth]{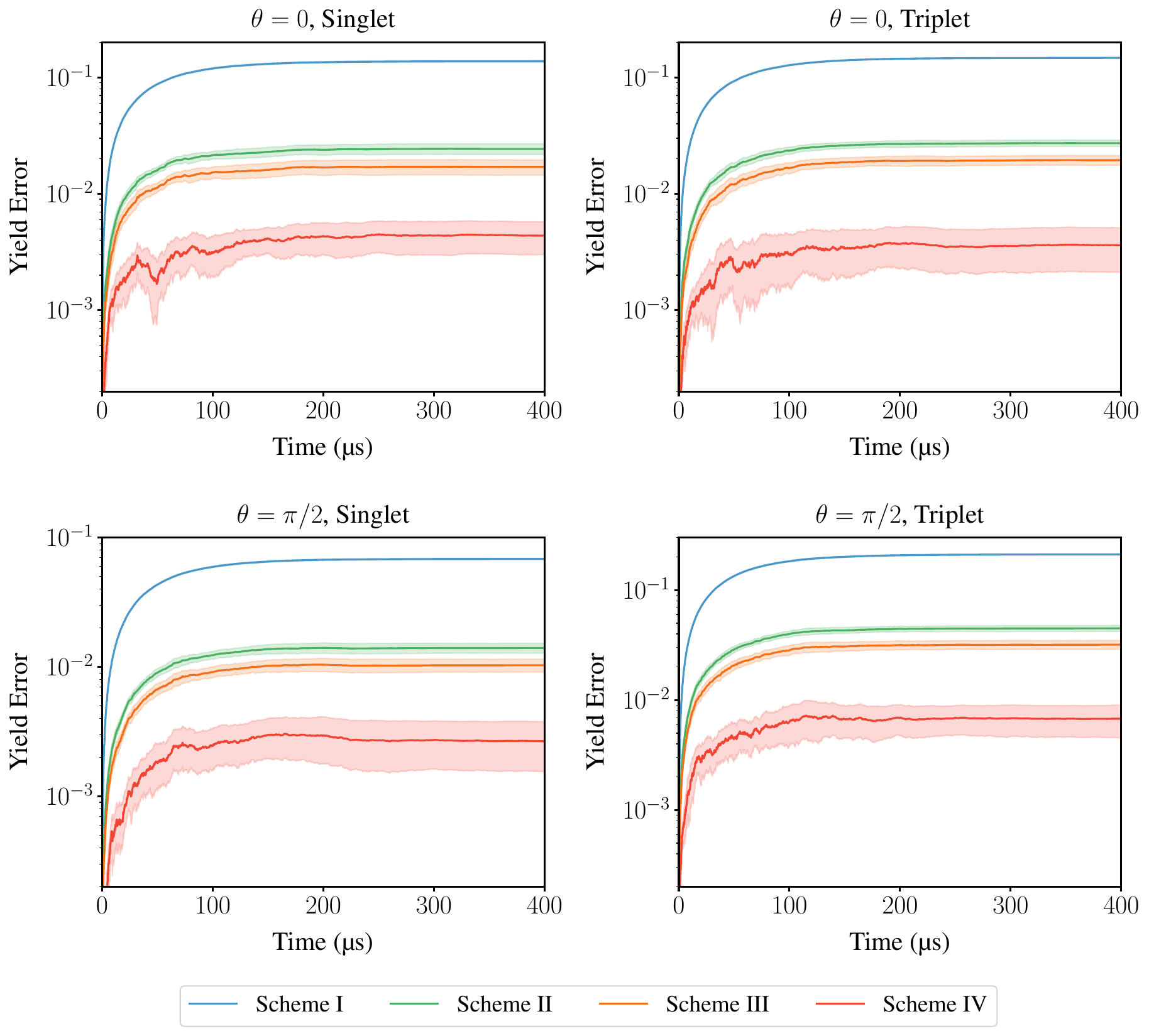}
    \caption{\raggedright Numerical tests of the Schemes I to IV using the RPM system. We set the angle to $0$ and $\pi/2$ and track the evolution of $\Tr(\rho \dyad{S})$ (i.e. singlet yield) and $\Tr(\rho \dyad{T})$ (i.e. triplet yield) for $\rho \approx \frac{1}{N_{\text{traj}}}\sum_{\ell =1}^{N_{\text{traj}}} \psi_\ell\psi_\ell^\dagger$ with each scheme and the same step size $\Delta = 1\times 10^{-7}$ for all the simulations up to the stopping time $400$ \textmu s. In each test, we adopt $N_{\text{traj}} = 10^4$. We compare the error of each scheme throughout the dynamics. The $y$ axis is plotted in the logarithmic scale. We repeat the experiment $20$ times with different random seeds, and the shaded area represents the 99\% confidence interval obtained from this sample.}
    \label{fig:RPM_each_order}
\end{figure}

There are also some additional structures in RPM that are helpful for the derivation of each scheme. RPM possesses the properties that
\begin{equation}\label{eq:orthogonality}
    \mathrm{Span}\left(\ket{S},\ket{T_0},\ket{T_+},\ket {T_-}\right) \perp  \mathrm{Span}\left(\ket{s,\uparrow},\ket{s,\downarrow},\ket{t_0, \uparrow},\ket{t_0,\downarrow},\ket{t_+, \uparrow},\ket{t_+,\downarrow},\ket{t_-, \uparrow},\ket{t_-,\downarrow}\right).
\end{equation}
Therefore, for any pair of jump operators $L_i, L_j$ ($i,j=1,2,\cdots,8$), we have $L_iL_j =L_jL_i= 0$ and thus $[L_i,L_j]=0$. In other words, in terms of the notations in \cref{eq:stomagnus}, we have $[G_i, G_j] = 0$ for any $i,j\ge 1$. This commutation property allows us to focus solely on the coefficients of the commutators $[G_0,G_j]$ ($j\ge 1$) in Scheme II, and the coefficients of the nested commutators $[G_0,[G_j,G_0]]$, $\Big[\big[[G_j, G_0],G_0\big],G_0\Big]$ ($j\ge 1$) in Schemes III and IV. This significantly simplifies the expressions of high-order schemes. We provide an illustrative demonstration of the each order scheme and present the explicit form of the Magnus integrators $\Omega^{[i]}$ ($i=1,2,3,4$) in \cref{appD}.

\begin{figure}[htbp]
    \centering
    \includegraphics[width=0.55\linewidth]{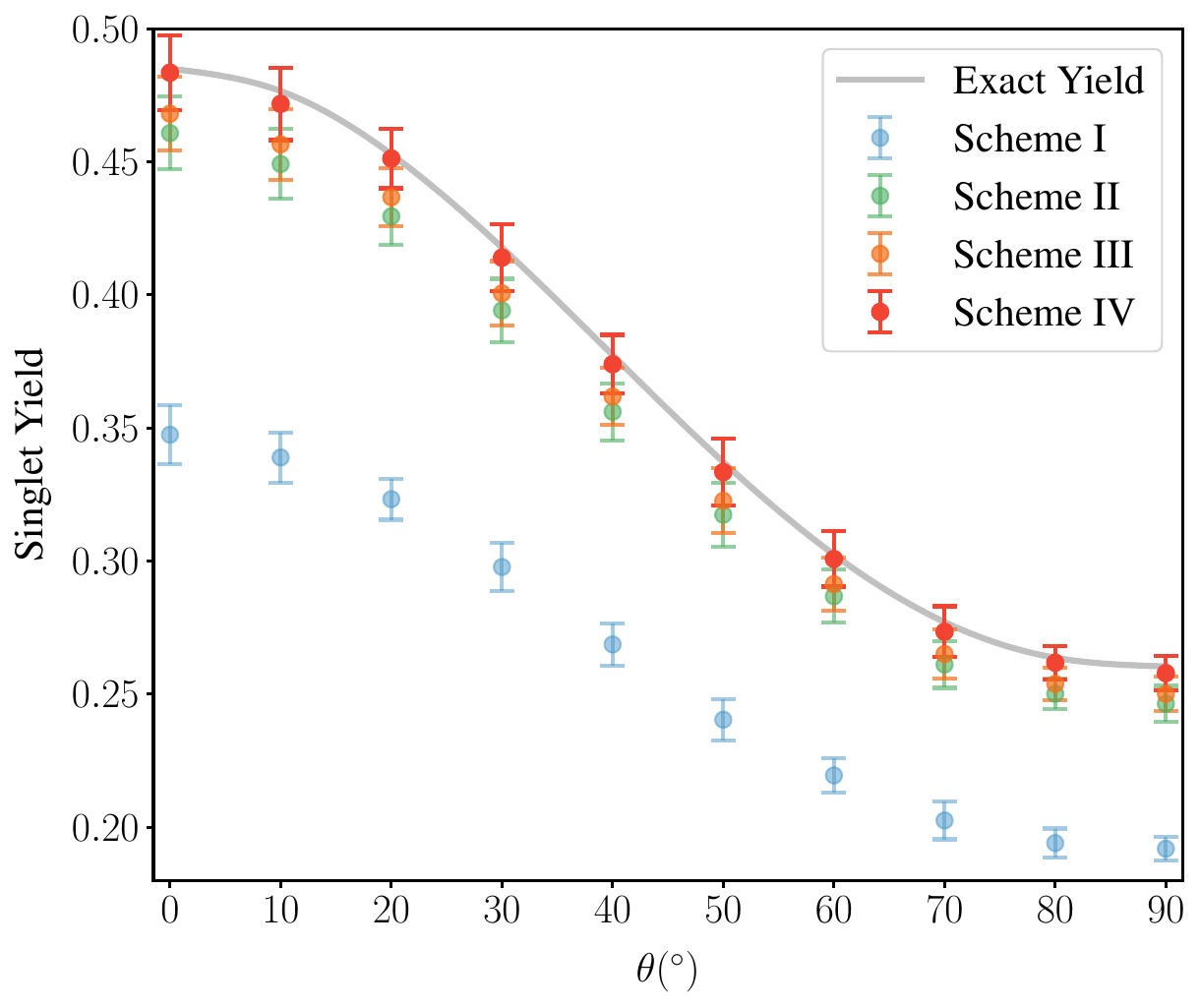}
    \caption{\raggedright Numerical tests of the Schemes I to IV for calculating the steady state properties using the RPM system. The density operators $\rho_\star$ at the stopping time $400$ \textmu s calculated using each scheme are considered as the approximation of the exact steady state. We compare the final singlet yield $\Tr(\rho_\star\dyad{S})$(or the sensitivity) versus the angle $\theta$ for $\theta = 0^\circ, \cdots, 90^\circ$ obtained using each order scheme. We repeat the experiment $20$ times with different random seeds, and the error bar represents the 99\% confidence interval.}
    \label{fig:RPM_angle_yields}
\end{figure}

The initial state is chosen as $ \rho_0 = \frac{1}{2} I \otimes \ket{s} \bra{s} $, and the simulation is performed using Schemes I to IV up to $ T = 400 $ \textmu s. Each simulation incorporates $ 10^4 $ trajectories, and the numerical experiment is repeated 20 times to ensure statistical reliability. We analyze the evolution of the error in the overlaps with the singlet state $ \ket{S} $ and the triplet state $ \ket{T} $ across different scheme orders. As illustrated in \cref{fig:RPM_each_order}, the error systematically decreases from Scheme I to Scheme IV. Notably, the Scheme IV exhibits excellent agreement with the exact solution for both singlet and triplet yield as well as $ \theta = 0^\circ $ and $ \theta = 90^\circ $, demonstrating its superior accuracy.  

To further test the robustness, we examine the steady state properties of the Lindblad dynamics. Specifically, we consider the sensitivity of the avian compass versus the angle $ \theta $. Here the sensitivity is characterized by the final singlet yield i.e. the overlap between the singlet state and the steady state of the Lindbladian. In our simulations, the density operator at the stopping time is taken as an approximation of the exact steady-state solution $ \rho_\star $, and we evaluate the overlap $ \Tr(\rho_\star \dyad{S}) $ accordingly. As shown in \cref{fig:RPM_angle_yields}, Scheme IV demonstrates significantly improved agreement with the exact solution in capturing the sensitivity across all considered angles $ \theta = 0^\circ, 10^\circ, \dots, 90^\circ $.
\section{Discussion and outlook}

To the best of our knowledge, this paper presents the \emph{first} variational quantum simulation algorithm for Lindblad equations with \emph{high-order precision}. We employ stochastic Magnus expansion to explicitly derive the single-step method of arbitrarily high order for solving the unraveled Lindblad dynamics. This method, formulated as an exponential integrator, is well-suited for NISQ-friendly variational implementation. Compared to previous methods, our wavefunction trajectory-based approach circumvents the formidable dimensionality challenge \cite{SchlimgenHeadMarsdenSagerEtAl2021,SchlimgenHeadMarsdenSagerEtAl2022,HuXiaKais2020, HuHeadMarsdenMazziottiEtAl2022, ZhangHuWangKais2023,HannLeeGirvinJiang2021,PeetzSmartTserkisNarang2024,KamakariSunMottaMinnich2022,SchlimgenHeadMarsdenSagerSmithEtAl2022,OhKrogmeierSchlimgenEtAl2024}, while the incorporation of high-order schemes enhances simulation robustness and accuracy \cite{LuoLinGao2024, LanLiang2024}, even in highly oscillatory regimes such as RPM.

Our new algorithm bridges the gap between the variational quantum simulations of Lindblad equation with the high-precision numerical SDE solvers. Moreover, this work introduces a new class of quantum dynamics problem, characterized by the presence of nonlinear terms and multiple groups of stochastic processes within the SDE formalism. The identification of suitable classical and quantum simulation methodologies presents a compelling direction for future research. For instance, we may introduce an appropriate change of variables based on the interaction-picture framework, also known as the right correction Magnus series (RCMS)~\cite{DeganiSchiff2006}, thereby further enhancing the efficiency and accuracy of simulations for nonlinear and highly oscillatory SDEs in quantum dynamics.

Compared to the quantum simulation of unitary dynamics, where a variety of algorithms have been extensively developed for both near-term and fault-tolerant quantum computing, the simulation of Lindblad dynamics and more general non-Hermitian quantum processes, such as non-Markovian open quantum system dynamics, remains in its early stages of research. This work introduces a novel framework to the field, and the Schemes III and IV may already be practical effective. Furthermore, we hope this study has the potential to advance the development of methodologies for simulating complicated quantum systems using realistic quantum devices, coinciding with the 100th anniversary since the initial development of quantum mechanics.
\begin{acknowledgments}
This work is financially supported by the National
Natural Science Foundation of China (NSFC) under Grant Nos.
22222605 and 223B1011, as well as the National Key Research and Development Project under Grant No.2022YFA1503900. The Tsinghua Xuetang Talents Program and High-Performance Computing Center of Tsinghua University were acknowledged for providing computational resources. 
\end{acknowledgments}

\bibliography{main}

\onecolumngrid
\clearpage

% \begin{center}
%     \textbf{Supplementary Materials}
% \end{center}

\appendix
\section{Norm variations in nonlinear and linear QSD}\label{appA}

In the generic nonlinear QSD \cref{eq:nlqsd}, the change in norm can be evaluated as
\begin{equation}
\begin{aligned}
\mathrm{d} |\psi|^2&=\mathrm{d}\text{Tr}(\psi \psi^\dagger)\\
&= \text{Tr}\left((\mathrm{d}\psi)\psi^\dagger+\psi(\mathrm{d}\psi)^\dagger+(\mathrm{d}\psi)(\mathrm{d}\psi)^\dagger\right)\\
&=\text{Tr}\left(\mathcal{L}[\rho_\psi] \dd t+ \sum_k(L_k-\langle L_k\rangle)\rho_\psi \mathrm{d}W_t^k + \sum_k\rho_\psi(L_k^\dagger-\langle L_k^\dagger\rangle) \mathrm{d}W_t^{k^*}\right)\\
& = 0,
\end{aligned}
\end{equation}
where we use the rule $\mathrm{d}W^2=\mathrm{d}t$ according to Itô calculus and ignore the deterministic terms with order higher than $\mathcal{O}(\mathrm{d}t^2)$. Here $\rho_\psi$ denote the initial density operator $\psi\psi^\dagger$, and the Lindblad evolution is trace-preserving, which implies $\Tr(\mathcal{L}[\rho_\psi]\dd t)=\Tr(\dd \rho_\psi)=0$. 

For linear QSD \cref{eq:linear}, it can also be derived that
\begin{equation}
\begin{aligned}
   \mathrm{d} |\psi|^2&=\text{Tr}\left(\sum_kL_k\rho_\psi \mathrm{d}W_t^k + \sum_k\rho_\psi L_k^\dagger \mathrm{d}W_t^{k^*}\right)\\
   &=\sum_k\langle L_k \rangle \mathrm{d}W_t^k + \sum_k \langle L_k^\dagger \rangle \mathrm{d}W_t^{k^*} \neq 0, 
\end{aligned}
\end{equation}
and the weight of single trajectory is given by the changing norm. However, the discretization of time still introduces changes in the norm for nonlinear quantum state diffusion (QSD), leading to non-Hermitian dynamics that are disturbed by certain Gaussian noise. In our work, the propagation of state is implemented on quantum circuits, and the norm remains unit.

\section{Understanding the conceptual distinctions between stochastic Magnus and Taylor expansions}\label{app:magnus_taylor}

In this section, we provide a conceptual comparison between the stochastic Magnus expansion and the stochastic Taylor expansion \cite{KloedenPlaten2013}. While both of these two methods can be used to derive the high-order numerical schemes for solving the QSD equation, the derivation and implementation of high-order schemes based on the stochastic Taylor expansion can be rather intricate, especially for higher order expansion and multiple dimension SDEs. 

Specifically, for general multi-dimensional SDEs, the stochastic Taylor integrator admits the form
    \begin{equation}
        Y_{k+1} = \sum_{\alpha\in \mathcal{A}_\gamma } \underline{f}_\alpha (t_k, Y_k)J_\alpha,\quad \mathcal{A}_\gamma =\{\alpha \in \mathcal{M}\mid l(\alpha)+n(\alpha)\leq 2\gamma\},
    \end{equation}
 where $\mathcal{M}$ denotes the set of all multi-indices, $l(\alpha)$ is the length of $\alpha$, $n(\alpha)$ counts the number of zeros in $\alpha$, and $\gamma$ specifies the order of the integrator. The terms $J_\alpha$ are iterated Stratonovich integrals, and the coefficient functions $\underline{f}_\alpha$ are defined recursively by
    \begin{equation}
        \underline{f}_\alpha = \left\{\begin{aligned}
            f \qquad \quad &l(\alpha)=0,\\
            \underline{L}^{j_1}\underline{f}_{-\alpha}\quad &l(\alpha)\geq 1,
        \end{aligned}\right.
    \end{equation}
where $\underline{L}^j=\sum_{k=1}^d b^{k,j}\frac{\partial }{\partial x^k}$ and $-\alpha$ denotes the multi-index obtained by removing the first element of $\alpha$. 

The recursive nature of the coefficient construction, together with the combinatorial complexity of the multi-index set $\mathcal{A}_\gamma$, renders the stochastic Taylor expansion increasingly cumbersome at higher orders. By contrast, the stochastic Magnus expansion offers a more algebraically structured framework. %in which the evolution operator is represented as a time-ordered exponential involving nested commutators. This structure not only facilitates a more compact representation of the solution but also preserves essential geometric and algebraic properties of the underlying dynamics, particularly in the context of non-commutative operator-valued SDEs.

\section{Derivation of Stratonovich-type linear and nonlinear QSD equations}\label{appB}
In this section, we provide a sketch of the results of \citet{Kunita1980} to demonstrate the transformation of \cref{eq:nlqsd,eq:linear} into Stratonovich-type SDEs. 

For a general Itô-type SDE
\begin{equation}
    \mathrm{d}\mathbf{X}_t = a(\mathbf{X}_t,t)\mathrm{d}t+b(\mathbf{X}_t,t)\mathrm{d}\mathbf{W}_t
\end{equation}
where $\mathbf{X}_t\in \mathbb{C}^n$ is a $n$-dimensional random variable and $\mathbf{W}_t\in\mathbb{C}^m$, the corresponding Stratonovich-type SDE can be obtained by the transformation
\begin{equation}
a^i(\mathbf{X}_t,t)\mapsto  \underline{a}^i(\mathbf{X}_t,t)\equiv a^i(\mathbf{X}_t,t)-\frac{1}{2}\sum_{k=1}^m\sum_{j=1}^n\frac{\partial b^{i,k}(\mathbf{X}_t,t)}{\partial \mathbf{X}_t^j}b^{j,k}(\mathbf{X}_t,t) ,\quad b(\mathbf{X}_t,t)\mapsto \overline b(\bv X_t, t) \equiv b(\mathbf{X}_t,t).
\end{equation}
In this way, we have $\bv X_t$ also solves the equivalent Stratonovich-type SDE:
\begin{equation}
    \dd \bv X_t = \underline a(\bv X_t, t) \dd t + b(\bv X_t , t ) \circ \dd \bv W_t,
\end{equation}
where the small circle $\circ$ before the Brownian differential denotes the Stratonovich format.

We begin with the derivation of the Stratonovich-type linear QSD equation \cref{eq:linear}. 
We assume that there are $m$ jump operators $\{L_k\}_{k=1}^m$, hence
\begin{equation}
\underline a(\psi_t,t) = -\ii H_{\rm eff} \psi_t -\frac12\sum_{k=1}^m L_k^2 \psi_t = \ii H_s \psi_t-\frac12 \sum_{k=1}^m (L_k^\dagger +L_k) L_k \psi_t,
 \end{equation}
\begin{equation}
    \underline b(\psi_t,t) = b(\psi_t,t) = (L_1\psi_t,L_2\psi_t,\dots, L_m\psi_t) \in \mathbb C^{n\times m}.
\end{equation}

Hence we obtain the following linear Stratonovich-type QSD equation
\begin{equation}
      \dd \psi = -\ii H_s \psi \dd t -\frac12 \sum_k (L_k+L_k^\dagger)L_k \psi \dd t + \sum_k L_k \psi \circ \dd W_t.
\end{equation}

For the nonlinear QSD \cref{eq:nlqsd}, we have
\begin{equation}
    a(\psi_t,t) =  -\ii 
{H}_s\psi_t+\sum_{k=1}^m(\langle{L}_k^{\dagger}\rangle {L}_k-\frac{1}{2} {L}_k^{\dagger} {L}_k -\frac12 \langle L_k^\dagger \rangle \langle L_k\rangle )\psi_t, 
\end{equation}
\begin{equation}
b(\psi_t,t)=\left((L_1-\langle L_1\rangle)\psi_t, (L_2-\langle L_2\rangle)\psi_t,\dots, (L_m-\langle L_m \rangle)\psi_t \right) \in\mathbb{C}^{n\times m}.
\end{equation}

%In practice, since we are focusing on single-step evolution, we may assume that $\langle L_k\rangle$ and $\langle L_k^\dagger\rangle$ are constant with respect to $\psi_t$. Under this approximation, we compute $\underline{a}$ as follows:

Now we could compute $\underline{a}$ as follows
%\begin{equation}
%\begin{aligned}
%    \underline{a}^i(\psi_t,t)&={a}^i(\psi_t,t)-\frac{1}{2}\sum_{k=1}^m\sum_{j=1}^n\frac{\partial \left(\sum_{u=1}^nL_k^{i,u}\psi_t^u-\langle L_k\rangle \psi_t^i\right)}{\partial \psi^j_t}(L_k\psi_t-\langle L_k\rangle \psi_t)^j\\
%    & = a^i(\psi_t,t) - \frac12\sum_{k=1}^m \sum_{j=1}^n [L_k^{i,j}  -\delta_{ij}\langle L_k\rangle] (L_k\psi_t-\langle L_k\rangle \psi_t)^j  \\
%    &={a}^i(\psi_t,t)-\frac{1}{2}\sum_{k=1}^m(L_k^2\psi_t-2\langle L_k\rangle L_k \psi_t + \langle L_k\rangle ^2\psi_t)^i.
%    \end{aligned}
%\end{equation}
\begin{equation}
\begin{aligned}
    \underline{a}^i(\psi_t,t)&={a}^i(\psi_t,t)-\frac{1}{2}\sum_{k=1}^m\sum_{j=1}^n\frac{\partial \left(\sum_{u=1}^nL_k^{i,u}\psi_t^u-\langle L_k\rangle \psi_t^i\right)}{\partial \psi^j_t}(L_k\psi_t-\langle L_k\rangle \psi_t)^j\\
    &={a}^i(\psi_t,t)-\frac{1}{2}\sum_{k=1}^m\sum_{j=1}^n\frac{\partial \left(\sum_{u=1}^nL_k^{i,u}\psi_t^u-\sum_{x=1}^n\sum_{y=1}^n \psi_t^{x^*}  L_k^{x,y} \psi_t^y \psi_t^i\right)}{\partial \psi^j_t}(L_k\psi_t-\langle L_k\rangle \psi_t)^j\\
   & = a^i(\psi_t,t) - \frac12\sum_{k=1}^m \sum_{j=1}^n \left( L_k^{i,j}  -\sum_{x=1}^n \psi_t^{x^*}L_k^{x,j}\psi_{t}^i - \delta_{ij}\langle L_k\rangle \right) (L_k\psi_t-\langle L_k\rangle \psi_t)^j  \\
    &={a}^i(\psi_t,t)-\frac{1}{2}\sum_{k=1}^m\left(L_k^2\psi_t-2\langle L_k\rangle L_k \psi_t + (2\langle L_k\rangle ^2-\langle L_k^2 \rangle )\psi_t \right)^i,
    \end{aligned}
\end{equation}
where the notation $\psi^i$ denotes the $i$-th entry of the vector $\psi\in \mathbb C^n$ and $A^{i,j}$ denotes the $(i,j)$-entry of the matrix $A\in \mathbb C^{n\times m}$. Here the property $\pdv{\psi_t^{i^*}}{\psi_t^{j}}=0$ derived from Cauchy–Riemann equations is used. Therefore,
\begin{equation}
\begin{aligned}
    \underline{a}(\psi_t,t)&={a}(\psi_t,t)-\frac{1}{2}\sum_{k=1}^m(L_k^2- 2\langle L_k\rangle L_k + 2\langle L_k\rangle^2 - \langle L_k^2\rangle )\psi_t\\
    & = -\ii {H}_s\psi_t+\sum_{k=1}^m \left[\left(2\Re\langle L_k\rangle L_k -\frac12 ((L_k + L_k^\dagger )L_k )\right)\psi_t + c_{k,\psi}\psi_t\right],
    \end{aligned}
\end{equation}
where
\begin{equation}
    c_{k,\psi}=-\frac12 \langle L_k^\dagger \rangle \langle L_k\rangle -\langle L_k\rangle^2+\frac12 \langle L_k^2 \rangle.
\end{equation}
Therefore, the nonlinear stratonovich-type QSD equation is
\begin{equation}
    \dd\psi = -\ii {H}_s\psi \dd t+ \sum_k \left[2\Re (\langle L_k\rangle) L_k - \frac12 (L_k + L_k^\dagger )L_k + c_{k,\psi}\right]\psi \dd t + \sum_k ({L}_k - \langle{L}_k\rangle)\psi \circ \mathrm{d}W_{t}^k.
\end{equation}

We note that in differential equations, terms like $c_{k,\psi} \psi_t\dd t$ or $\langle L_k\rangle \psi_t\circ \dd W_t^k$ $(j=1,\cdots,n)$, where $c_{k,\psi},\langle L_k\rangle$ are complex numbers, only introduce a factor like $\exp[\sum_k\int  (c_{k,\psi} \dd t-\langle L_k\rangle \circ \dd  W_t^k) ]$ in the solution. These terms does not alter the structure of the equation except for introducing a scaling, which is irrelevant in the nonlinear QSD since the norm of state is preserved, as proved in \cref{appA}. Therefore, in practice, we can equivalently disregard these terms and evolve only the remaining part, while apply normalization at each step~\cite{Percival1998}. Then with some abuse of notation, this amounts to design the single step method for the following QSD equation up to the normalization at each step
\begin{equation}
  \dd\psi = -\ii {H}_s\psi \dd t+ \sum_k \left[2\Re (\langle L_k\rangle) L_k - \frac12 (L_k + L_k^\dagger )L_k \right]\psi \dd t + \sum_k {L}_k \psi \circ \mathrm{d}W_{t}^k.
\end{equation}

It is worth noting that on quantum circuits, normalization of the state is automatically ensured. Therefore, in VQS implementation we only need to update the parameters based on the variational principle in each step and no additional normalization operation is required.

\section{Stochastic Magnus expansion and multiple Stratonovich integrals}\label{appC}

As in the main text, we begin with the following general multi-dimensional Stratonovich-type SDE
\begin{equation}\label{eq:multidimsdeapp}
    \dd \bv X_t  = G_0 \bv X_t \dd t +\sum_{j=1}^d   G_j\bv X_t\circ \dd W^j_t .
\end{equation}
Moreover, we define formally that
\begin{equation}\label{eq:formalAtAt}
    A(t)\dd t =  G_0\dd t + \sum_{j=1}^d G_j \circ \dd  W^j_t.
\end{equation}
Then we can plug this into deterministic Magnus expansion \cref{eq:determagnus}, and after considerable derivations \cite{Burrage1999, BurrageBurrage1999}, it can be shown that the solution to the SDE \cref{eq:multidimsdeapp} is given by 
\begin{equation}
    \bv X_t = \exp(\Omega(t)) \bv X_0,
\end{equation}
where $\Omega(t)$ has been given in \cref{eq:stomagnus}. The time-dependent Stratonovich multiple integral is defined as
\begin{equation}
    J_{j_1j_2\dots j_k,t}\equiv \int_{0}^t\cdots \int_{0}^{s_3}\int_0^{s_2}\circ\, \dd W_{s_1}^{j_1}\circ \dd W_{s_2}^{j_2}\cdots \circ \dd W_{s_k}^{j_k},\quad \dd W_{t}^{0}\equiv \dd t.
\end{equation}
We provide details about how to evaluate these stochastic integrals in \cref{appD}.

We next demonstrate how to derive the explicit expression of the Scheme IV Magnus integrator $\Omega^{[4]}$. Recall that 
\begin{equation}
\begin{aligned}
    \Omega^{[4]}(t) &= \frac{1}{12}\int_0^t \dd s_1\int_0^{s_1}\dd s_2 \int_0^{s_2} \dd s_3 \int_0^{s_3} \dd s_4 \Bigg(\Big[\big[[A(s_1),A(s_2)],A(s_3)\big],A(s_4)\Big]  \\
    & + \Big[A(s_1),\big[[A(s_2),A(s_3)],A(s_4)\big]\Big] +\Big[A(s_1),\big[A(s_2),[A(s_3),A(s_4)]\big]\Big] +\Big[A(s_2),\big[A(s_3),[A(s_4),A(s_1)]\big]\Big] \Bigg ).
\end{aligned}
\end{equation}
For the sake of notation simplicity, we define
\begin{equation}\label{eq:Iijkl}
    I_{ijkl}(t) \equiv \int_0^t \dd s_1\int_0^{s_1}\dd s_2 \int_0^{s_2} \dd s_3 \int_0^{s_3} \dd s_4  \Big[\big[[A(s_i),A(s_j)],A(s_k)\big],A(s_l)\Big],
\end{equation}
where $\{ i,j,k,l\}$ is a permutation of $\{1,2,3,4\}$. Therefore, using the properties of the commutator, we have
\begin{equation}\label{eq:omega4}
    \Omega^{[4]}(t)=\frac{1}{12}\left( I_{1234}-I_{2341}+I_{3421}+I_{4132}\right)
\end{equation}
Thus, we could now directly derive the fourth term in stochastic Magnus expansion. As a demonstration, we consider only the case of RPM, where the jump operators satisfy some commutative relations, as discussed in \cref{sec:rpm_numer}. We expand $A(s_i)$ and $A(s_j)$ formally according to \cref{eq:formalAt} (or \cref{eq:formalAtAt}), yielding
\begin{equation}
\begin{aligned}
    &\quad\,\, \Big[\big[[A(s_i),A(s_j)],A(s_k)\big],A(s_l)\Big]\dd s_1 \dd s_2 \dd s_3\dd s_4 \\
    &=  \Big[\big[[A(s_i),A(s_j)],A(s_k)\big],A(s_l)\Big]\dd s_i \dd s_j \dd s_k\dd s_l \\
    &= \Big[\big[[G_0\dd s_i+\sum_{m=1}^d G_m\circ \dd W_{s_i}^m,G_0\dd s_j+\sum_{m=1}^d G_m\circ \dd W_{s_j}^m],A(s_k)\big],A(s_l)\Big]\dd s_k\dd s_l \\
    &=\sum_{m=1}^d\Big[\big[[G_0, G_m],A(s_k)\big],A(s_l)\Big]\circ \dd s_i\dd W_{s_j}^m\dd s_k\dd s_l +\sum_{m=1}^d\Big[\big[[G_m, G_0],A(s_k)\big],A(s_l)\Big]\circ \dd W_{s_i}^m\dd s_j\dd s_k\dd s_l
\end{aligned}
\end{equation}
Note that for any $G_m$ in RPM, we can simplify each nested commutator as follows (see \cref{eq:orthogonality}):
\begin{equation}
    [G_m,G_0] = G_mG_0-G_0G_m=G_mG_0,
\end{equation}
\begin{equation}
    \big[ [G_m,G_0],G_n\big]=[G_mG_0,G_n]=0,
\end{equation}
\begin{equation}
    \Big[\big[[G_m,G_0],G_0\big],G_n\Big]=\big[[G_mG_0, G_0],G_n\big]=[G_mG_0^2,G_n]=0.
\end{equation}
Therefore,
\begin{equation}
    \Big[\big[[G_0, G_m],A(s_k)\big],A(s_l)\Big]\circ \dd s_i\dd W_{s_j}^m\dd s_k\dd s_l = - \Big[\big[[G_m, G_0],G_0\big],G_0\Big] \circ \dd s_i\dd W_{s_j}^m\dd s_k\dd s_l 
\end{equation}
\begin{equation}
    \Big[\big[[G_m, G_0],A(s_k)\big],A(s_l)\Big]\circ \dd W_{s_i}^m\dd s_j\dd s_k\dd s_l =  \Big[\big[[G_m, G_0],G_0\big],G_0\Big] \circ \dd W_{s_i}^m\dd s_j\dd s_k\dd s_l 
\end{equation}

We are now at the place to evaluate each term in \cref{eq:omega4}. From \cref{eq:Iijkl}, we first note that
\begin{equation}
    I_{1234}=\sum_{m=1}^d \Big[\big[[G_m, G_0],G_0\big],G_0\Big](J_{000m}-J_{00m0})
\end{equation}
By permuting the indices, we have 
\begin{equation}
    I_{2341} = \sum_{m=1}^d \Big[\big[[G_m, G_0],G_0\big],G_0\Big](J_{00m0}-J_{0m00}),
\end{equation}
\begin{equation}
    I_{3421} = \sum_{m=1}^d \Big[\big[[G_m, G_0],G_0\big],G_0\Big](J_{0m00}-J_{m000}),
\end{equation}
\begin{equation}
    I_{4132} = \sum_{m=1}^d \Big[\big[[G_m, G_0],G_0\big],G_0\Big](J_{m000}-J_{000m}).
\end{equation}
Thus in RPM, we arrive at
\begin{equation}
    \Omega^{[4]}(t)=\Omega^{[3]}(t)+\frac{1}{6}\sum_{m=1}^d \Big[\big[[G_m, G_0],G_0\big],G_0\Big](J_{0m00}-J_{00m0}).
\end{equation}

\section{Illustrative derivation of multiple Stratonovich integrals}\label{appD}

There is a systematic approach to represent the multiple Stratonovich Integrals. The derivation details of \cref{eq:fourierJjt} are fully described in \citet{KloedenPlaten2013} and we only recite part of these results. 

In the forthcoming discussion, we assume $j,j_1,j_2\in\{1,2,\dots,m\}$. Applying the Fourier expansion to the Brownian bridge process $\{ W_t-\frac{t}{\Delta}W_\Delta, 0\leq t\leq \Delta \}$ on the time interval $[0, \Delta]$, we have 
\begin{equation}\label{eq:fourierJjt}
    J_{j,t}=\frac{W_\Delta^j}{\Delta}t+\frac{1}{2} a_{j,0}+\sum_{r=1}^\infty \left( a_{j,r}\cos{\frac{2\pi rt}{\Delta}}+b_{j,r}\sin{\frac{2\pi rt}{\Delta}}\right).
\end{equation}
Here, $\{W^j_\Delta\}$ are a set of independent Gaussian random variables with zero mean and variance $\Delta$, $a_{j,0} \sim \mc N\left(0,\frac \Delta 3\right)$, and the pairwise independent stochastic coefficients $a_{j,r}, b_{j,r}$ are $\mathcal{N}(0,\frac{\Delta}{2\pi^2r^2})$ distributed. 

For integrals with multi-indices of length $1$ or $2$, we have
\begin{equation}\label{eq:stochasticintegral1}
    J_0 = \Delta,\quad J_j = W_\Delta^j
\end{equation}
and
\begin{equation}
    \quad J_{00}=\frac{1}{2}\Delta^2,\quad J_{0j}=\frac{1}{2}\Delta(W_\Delta^j-a_{j,0}),\quad J_{j0}=\frac{1}{2}\Delta(W_\Delta^j+a_{j,0}),\quad 
\end{equation}
\begin{equation}
    J_{j_1j_2}=\frac12 W_\Delta^{j_1} W_\Delta^{j_2}-\frac12 (a_{j_2,0}W_\Delta^{j_1}-a_{j_1,0}W_\Delta^{j_2})+\Delta A_{j_1,j_2}
\end{equation}
respectively, where
\begin{equation}
    A_{j_1,j_2}=\frac{\pi}{\Delta}\sum_{r=1}^{\infty}r(a_{j_1,r}b_{j_2,r}-b_{j_1,r}a_{j_2,r}).
\end{equation}
For integrals with multi-indices of greater length, where at most one index is nonzero, we have
\begin{equation}
J_{000}= \frac{1}{6}\Delta^3,\quad J_{0j0} =\frac{1}{6}\Delta^2W_\Delta^j-\frac{1}{\pi}\Delta^2b_j,
\end{equation}
\begin{equation}
J_{j00}=\frac{1}{6} \Delta^2 W_{\Delta}^j+\frac{1}{4}\Delta^2 a_{j,0}+\frac{1}{2 \pi} \Delta^2 b_j, \quad
J_{00j}=\frac{1}{6} \Delta^2 W_{\Delta}^j-\frac{1}{4} \Delta^2 a_{j,0}+\frac{1}{2 \pi} \Delta^2 b_j ,
\end{equation}
and
\begin{equation}
    b_j=\sum_{r=1}^{\infty} r^{-1} b_{j, r}.
\end{equation}
For more complex stochastic integrals of the forms like $J_{0j_1j_2},J_{j_10j_2},\dots,J_{j_1j_2j_3}$, we refer the reader to \citet{KloedenPlaten2013}. 

With these above, we are ready to provide the detailed computational procedure for part of coefficients in stochastic Magnus expansion. For the Scheme I, it is straightforward to write down the corresponding terms through \cref{eq:stochasticintegral1}. For the coefficients of $[G_0,G_j]$ in Scheme II, we can derive that
\begin{equation}
    \frac{1}{2}(J_{j0}-J_{0j}) = \frac14\Delta(a_{j,0}+a_{j,0}) = \frac12\Delta a_{j,0} \sim \mc N\left(0, \frac{\Delta^3}{12}\right),
\end{equation}
and for $[G_i,G_j]\, (i\neq 0)$, we have
\begin{equation}
    \frac12 (J_{ji}-J_{ij})=\frac12 \left(a_{j,0}W_{\Delta}^i-a_{i,0}W_{\Delta}^j +2\Delta A_{j,i}\right).
\end{equation}
Similarly, for the coefficients of $[G_0,[G_j,G_0]]$,
\begin{equation}
\frac{1}{3}(J_{0j0}-J_{j00})+\frac{1}{12}J_0(J_{j0}-J_{0j})= -\frac{\Delta^2}{12}a_{j,0}-\frac{\Delta^2}{2\pi}b_j+\frac{\Delta^2}{12}a_{j,0}=-\frac{\Delta^2}{2\pi}b_j\sim\mathcal{N}\left(0,\frac{\Delta^5}{720}\right).
\end{equation}
Here we use the fact that the sum of a set of independent Gaussian random variables is still Gaussian, and 
\begin{equation}\label{eq:varbj}
    \mathrm{Var}(b_j) = \sum_{r=1}^\infty \frac{1}{r^2}\mathrm{Var}(b_{j,r})  = \frac \Delta {2\pi^2}\sum_{r=1}^\infty \frac{1}{r^4}  = \frac{\zeta(4)}{2\pi^2}\Delta = \frac{\pi^2 \Delta }{180}.
\end{equation}
Here $\zeta$ denotes the Riemann-Zeta function. 

To illustrate the derivation of stochastic coefficients for higher-order schemes, we need to evaluate $J_{0m00}-J_{00m0}$, which appears in the Scheme IV and is used in the numerical simulation of RPM, as \cref{appC} shows.  Utilizing the Fourier expansion \cref{eq:fourierJjt}, we could derive that 
\begin{equation}
\begin{aligned}
        J_{0m00}&=\int_0^\Delta \int_0^{s_1}\int_0^{s_2}  \int_0^{s_3}\circ \, \dd s_4 \dd W_{s_3}^m \dd s_2 \dd s_1 \\
        &=\int_0^\Delta \int_0^{s_1}\int_0^{s_2}  s_3 \circ \dd W_{s_3}^m \dd s_2 \dd s_1 \\
        &=\int_0^\Delta \int_0^{s_1}\int_0^{s_2}  s_3 \left( \frac{W_\Delta^{m}}{\Delta} + \sum_{r=1}^\infty \frac{2\pi r}{\Delta}\left(-a_{m,r}\sin{\frac{2\pi r s_3}{\Delta}} +b_{m,r}\cos{\frac{2\pi r s_3}{\Delta}}\right)\right) \dd s_3 \dd s_2 \dd s_1 \\
        &=\int_0^\Delta \int_0^{s_1}\Bigg[\frac{1}{2}s_2^2 \left( \frac{W_\Delta^{m}}{\Delta} + \sum_{r=1}^\infty \frac{2\pi r}{\Delta}\left(-a_{m,r}\sin{\frac{2\pi r s_2}{\Delta}} +b_{m,r}\cos{\frac{2\pi r s_2}{\Delta}}\right)\right)\\
        &\quad -\int_0^{s_2} \frac{1}{2}s_3^2 \sum_{r=1}^\infty\frac{4\pi^2 r^2}{\Delta^2} \left(-a_{m,r}\cos{\frac{2\pi r s_3}{\Delta}} -b_{m,r}\sin{\frac{2\pi r s_3}{\Delta}}\right)\dd s_3\Bigg]\dd s_2 \dd s_1,
\end{aligned}
\end{equation}
and
\begin{equation}
    \begin{aligned}
        J_{00m0} &= \int_0^\Delta \int_0^{s_1}\int_0^{s_2}  \int_0^{s_3}\circ \, \dd s_4 \dd s_3 \dd W_{s_2}^m \dd s_1 \\
        &= \int_0^\Delta \int_0^{s_1}\frac{1}{2}s_2^2 \circ \dd W_{s_2}^m \dd s_1 \\
        &=\int_0^\Delta \int_0^{s_1}\frac{1}{2}s_2^2 \left( \frac{W_\Delta^{m}}{\Delta} + \sum_{r=1}^\infty \frac{2\pi r}{\Delta}\left(-a_{m,r}\sin{\frac{2\pi r s_2}{\Delta}} +b_{m,r}\cos{\frac{2\pi r s_2}{\Delta}}\right)\right) \dd s_2 \dd s_1.
    \end{aligned}
\end{equation}
Note that for $r\in\mathbb N$, we have
\begin{equation}
    \int_0^{2\pi r}\int_0^{z}\int_0^{y} x^2\sin x\dd x\dd y \dd z = 0,\quad  \int_0^{2\pi r}\int_0^{z}\int_0^{y} x^2\cos x\dd x\dd y \dd z = -24\pi r,
\end{equation}
thus 
\begin{equation}
    \begin{aligned}
        J_{0m00}-J_{00m0} &= \sum_{r=1}^\infty \frac{2\pi^2 r^2}{\Delta^2} \int_0^\Delta \int_0^{s_1} \int_0^{s_2} s_3^2\left(a_{m,r}\cos{\frac{2\pi r s_3}{\Delta}} +b_{m,r}\sin{\frac{2\pi r s_3}{\Delta}}\right) \dd s_3\dd s_2\dd s_1\\
        %&=\sum_{r=1}^\infty \frac{2\pi^2 r^2}{\Delta^2} \int_0^\Delta \int_0^{s_1} \int_0^{s_2} s_3^2\left(a_{m,r}\cos{\frac{2\pi r s_3}{\Delta}} +b_{m,r}\sin{\frac{2\pi r s_3}{\Delta}}\right) \dd s_3\dd s_2\dd s_1\\
        &=-\sum_{r=1}^\infty\frac{3\Delta^3}{2\pi^2r^2}a_{m,r}.
    \end{aligned}
\end{equation}
Recall that $\{a_{m,r}\}_{r\ge 1}$ are mutually independent and $a_{m,r}\overset{\mathrm{I.I.D.}}{\sim}\mathcal{N}(0,\frac{\Delta}{2\pi^2r^2})$ (see \cref{appC}). Therefore, similar to \cref{eq:varbj}, we have
\begin{equation}
    \mathrm{Var}\left(\frac16 (J_{0m00}-J_{00m0})\right) = \frac{\Delta^6}{16\pi^4}\sum_{r=1}^\infty\frac{1}{r^4}\mathrm{Var}(a_{m,r}) = \frac{\Delta^6}{32\pi^6}\sum_{r=1}^\infty\frac{1}{r^6}=\frac{\Delta^7\zeta(6)}{32\pi^6} = \frac{\Delta^7}{30240}.
\end{equation}
Then for the coefficients of $\Big[\big[[G_m, G_0],G_0\big],G_0\Big]$ in $\Omega^{[4]}$, we have
\begin{equation}
    \frac{1}{6} (J_{0m00}-J_{00m0}) \sim \mc N\left(0,\frac{\Delta^7}{30240}\right).
\end{equation}

In practice, we often truncate the infinite stochastic series in these integrals and retain only the first $p$ terms. The mean-square error between the approximation $J_\alpha^p$ and $J_\alpha$ where $\alpha = (j_1,j_2,\dots, j_k)$ can be estimated as~\cite{KloedenPlaten2013}
\begin{equation}
    \mathbb{E}(\abs{J_\alpha^p-J_\alpha}^2)\leq \frac{\Delta^2}{2\pi^2 p}.
\end{equation}
Fortunately, under most situations, computing higher-order integrals is not strictly necessary, as accuracy of Scheme II is sufficient.

\section{Details of variational quantum simulation for the QSD dynamics}\label{appE}

\begin{figure}[!h]
    \subcaptionbox{\label{fig:CirCuit}}{\includegraphics{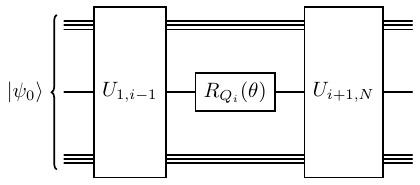}}
    \subcaptionbox{\label{fig:CirCuitDiff}}{\includegraphics{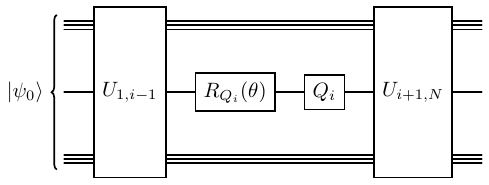}}
    \caption{\label{fig:CirCuitAndDiff}The quantum circuits that represent the states $|\psi(\Theta(t))\rangle$ and $|\xi_i(\Theta(t))\rangle$, respectively.}
\end{figure}

Based on the parameter shift rule~\cite{MariaVilleIzaac2019}, the matrix $\mathbf{M}$ and vector $\mathbf{V}$ where
\begin{equation}
    M_{i,j} = \text{Re}\left\langle \frac{\partial \psi(\Theta (t))}{\partial \theta _i}\Bigg | \frac{\partial \psi(\Theta (t))}{\partial \theta _j}\right \rangle, \quad V_i= \text{Im} \left\langle \frac{\partial \psi(\Theta (t))}{\partial \theta _i} \right| \mc H \left | \psi(\Theta (t))\right \rangle
\end{equation}
can be directly measured from the quantum circuits. To reduce the measurement complexity for quantum devices, we restricted the parametrized quantum gates to single-qubit rotation gates $R_{Q}(\theta)$ in HVA ansatz such that
\begin{equation}
    R_{Q}(\theta)=e^{-\frac{\ii\theta}{2}Q},
\end{equation}
where $Q\in \{\sigma_x,\sigma_y,\sigma_z\}$ is a Pauli matrix. This can be achieved by decomposing the standard HVA ansatz into basic quantum gates and appropriately adjusting the variational circuit. For convenience, we write the ansatz state as
\begin{equation}
    |\psi(\Theta(t))\rangle = \prod_{j=1}^{N}e^{-\frac{\ii\theta_j(t)}{2}Q_j}|\psi_0\rangle,
\end{equation}
where non-parametrized quantum gates, for example, CNOT, are all omitted. Specifically, we now examine the differential of the $i$-th variable. As illustrated in \cref{fig:CirCuit}, we place the products of  quantum gates on the left and right of the corresponding rotation gate $R_{Q_i}(\theta_i)$, denoted as $U_{1,i-1}$ and $U_{i+1,N}$ respectively. Then the partial derivative of wave function ansatz can be expressed as
\begin{equation}
 \left |\pdv{\psi(\Theta(t))}{\theta_i} \right\rangle= \pdv{(U_{i+1,N} e^{-\frac{\ii\theta_i}{2}Q_i}U_{1,i-1})}{\theta_i}|\psi_0\rangle=-\frac{\ii}{2} U_{i+1,N} Q_i e^{-\frac{\ii\theta_i}{2}Q_i}  U_{1,i-1} |\psi_0\rangle\equiv -\frac{\ii}{2}|\xi_i(\Theta(t))\rangle.
\end{equation}

We see that the partial derivative of such state is simply obtained by adding a Pauli gate in front of (or behind, since they commute in this case) the corresponding parameterized rotation gate. Then we can directly obtain the entries of $\mathbf{M}$ and $\mathbf{V}$ via Hadamard test circuits, simulated using \texttt{Qiskit}\cite{Qiskit2024}. To be more specific, we illustrate how the matrix elements of $\bv M$ can be obtained from measurements on quantum circuits. The diagonal elements of $\bv M $ are $\frac{1}{4}$ since
        \begin{equation}
           \bv M_{i,i} = \text{Re}\left\langle \frac{\partial \psi(\Theta (t))}{\partial \theta _i}\Bigg | \frac{\partial \psi(\Theta (t))}{\partial \theta _i}\right \rangle = \frac{\langle\xi_i|\xi_i\rangle}{4}=\frac{1}{4}.
        \end{equation} 

\begin{figure}[!h]
    \subcaptionbox{\label{fig:HadamardTestE}}{\includegraphics{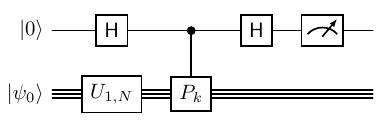}}
    
    \subcaptionbox{\label{fig:HadamardTestM}}{\includegraphics{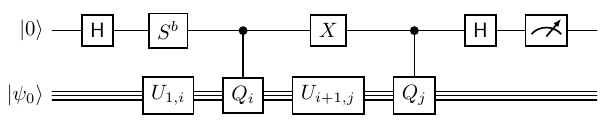}}
    \subcaptionbox{\label{fig:HadamardTestV}}{\includegraphics{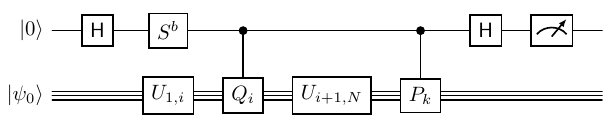}}

    \caption{\label{fig:HadamardTest}\raggedright The general Hadamard test quantum circuit to measure (a) $\langle \psi |P_k|\psi\rangle$; (b) $\langle \xi_i |\xi_j\rangle$; (c) $\langle \psi | P_k |\xi_i\rangle$. Here $\mathsf H$ denotes the Hadamard gate, $S$ denotes the phase gate and the binary integer $b\in \{0,1\}$ determines whether the measurement result corresponds to the real or imaginary part. Similar to \cref{fig:CirCuitAndDiff}, $U_{i,j}$ represents the set of parameterized quantum gates from the $i$-th to the $j$-th, and $Q_i,Q_j$ are Pauli gates. Here, $P_k$ denotes the Pauli string, which is obtained by performing the decomposition $\mathcal{H}=\sum c_kP_k$ on the operator to be measured.}
\end{figure}

To evaluate the off-diagonal element $\bv M_{i,j}$ with $i\neq j$ via the Hadamard test, we first prepare the following quantum state,
       \begin{equation}
            |\psi_{ij}\rangle =\frac{|0\rangle_{\text{anc}} \otimes |\xi_j\rangle + |1\rangle_{\text{anc}} \otimes |\xi_i\rangle}{\sqrt{2}},
        \end{equation}
        where ``anc" stands for the ancilla qubit. This preparation can be achieved by inserting a Controlled-$Q_k$ gate after the gate $R_{Q_k}(\theta)$. An $X$ gate on the ancillary qubit is inserted between the two controlled gates to invert the coupling to the ancillary qubit, as illustrated in \cref{fig:HadamardTestM}. The measurement of V is based on a similar procedure. In addition, we also need to measure the expectation value of jump operators $\{L_k\}$. The corresponding quantum circuit is illustrated in \cref{fig:HadamardTest}.

\begin{figure}[!h]
    \subcaptionbox{\label{fig:TFIM_CQ_Comp} A single trajectory evolution of the state populations in the TFIM with damping.}{\includegraphics[width=0.9\linewidth]{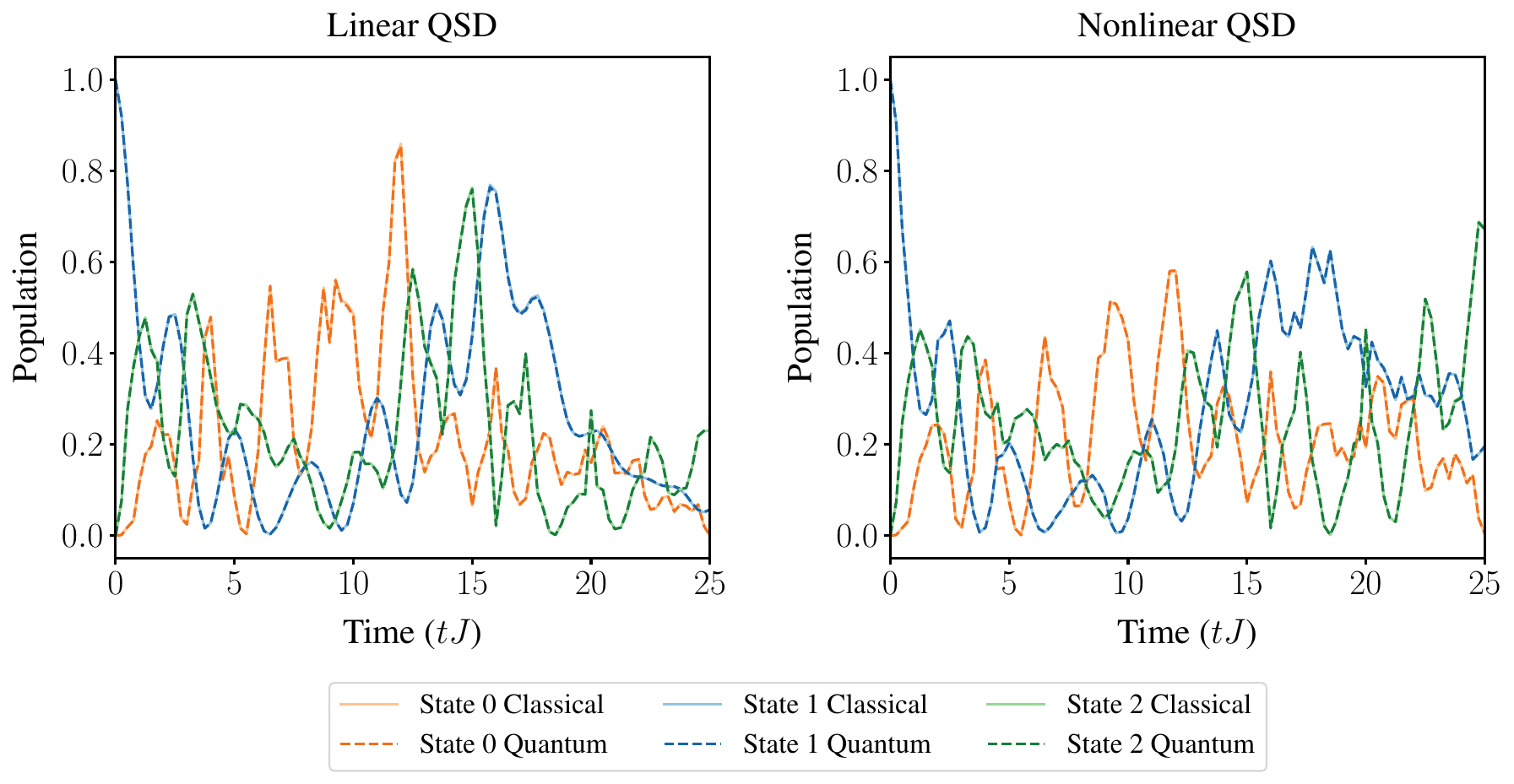}}\hspace{11mm}
    
    \subcaptionbox{\label{fig:FMO_CQ_Comp}A single trajectory evolution of the state populations in the FMO system.}{\includegraphics[width=0.9\linewidth]{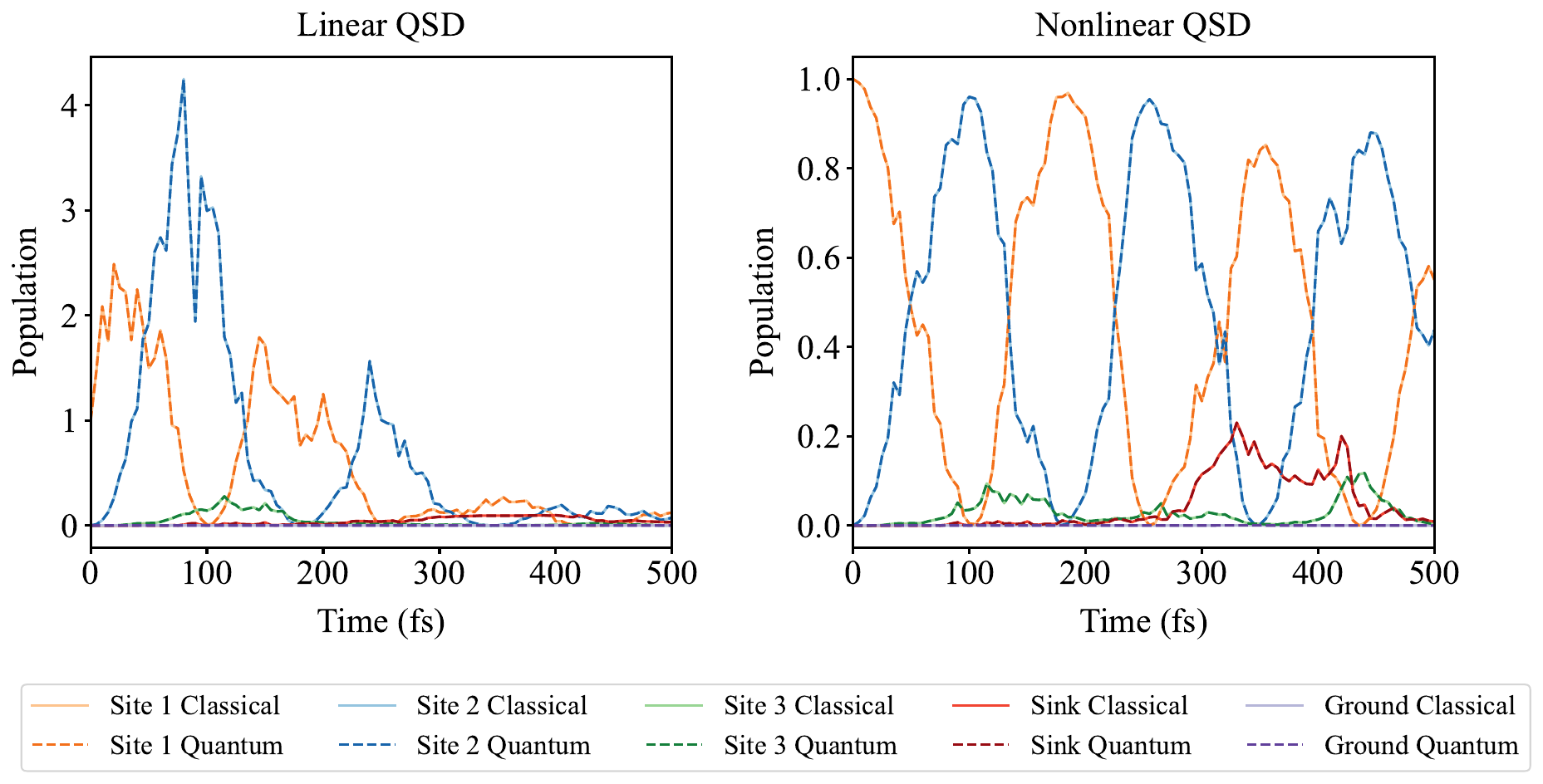}}

    \caption{\label{fig:CQ_Comp} \raggedright The comparison between the linear and nonlinear QSD evolution obtained by the exact wavefunction vector treatment (solid line) and the variational quantum time evolution (dashed line) with the same random seeds. For illustration, we employ the Scheme I Magnus integrator in each numerical simulation, which is implemented using \texttt{Qiskit}.}
\end{figure}
Note that we focus solely on ideal, noise-free qubits in this work. 
To mitigate computational costs, we choose to perform classical propagation directly via $\psi_{n+1}=\exp{(\Omega(\Delta))}\psi_n$ instead of performing variational simulations on the quantum circuit. The viability of this alternative approach has been verified at the single-trajectory level. \cref{fig:CQ_Comp} illustrates that the results obtained via variational simulations using \texttt{Qiskit}\cite{Qiskit2024} are fully consistent with those from direct classical evolution.

%To mitigate computational costs, we opt for direct classical computation instead of performing variational simulations on the quantum circuit. The viability of this alternative approach has been verified at the single-trajectory level—\cref{fig:CQ_Comp} illustrates that the results obtained via variational simulations using \texttt{Qiskit} \cite{Qiskit2024} are fully consistent with those from classical computations. This further demonstrates the feasibility of our algorithm. 

\section{Additional numerical comparison of Magnus integrator and Euler-Maruyama method}\label{app:EM_magnus_compare}

As sketched in \cref{sec:sketch_numerSDE}, the simplest numerical scheme for solving the QSD equation is the Euler-Maruyama method \cref{eq:EulerMaruyama}, which has a weak convergence order $\mc O(\Delta)$. We point out that even the Scheme I Magnus integrator performs significantly better than the Euler-Maruyama method, which is also consistent with the findings of some previous relevant studies \cite{LiLi2019}.

To facilitate an at-a-glance comparison, we conduct additional numerical experiments on the TFIM with damping (see \cref{sec:tfim}). Following the setup of the main text, we simulate $N_{\text{traj}} = 10^3$ trajectories with a final time of $T = 25~tJ$. For the Euler–Maruyama simulations, we adopt time steps $\Delta = 2.5 \times 10^{-3}$, $2.5\times 10^{-2}$, and $2.5\times 10^{-1}~tJ$, corresponding to $10^4$, $10^3$, and $10^2$ steps, respectively. The Magnus integrator simulations are carried out using a single time step size of $\Delta = 2.5\times 10^{-1}~tJ$ (i.e., $10^2$ steps). Each configuration is repeated 10 times with different random seeds, and we evaluate the mean overlap error with respect to the basis states $\ket{00}$, $\ket{01}$, and $\ket{11}$ at each step, with error bars denoting the 99\% confidence interval.

\begin{figure}[!h]
\includegraphics[width = 0.80\textwidth]{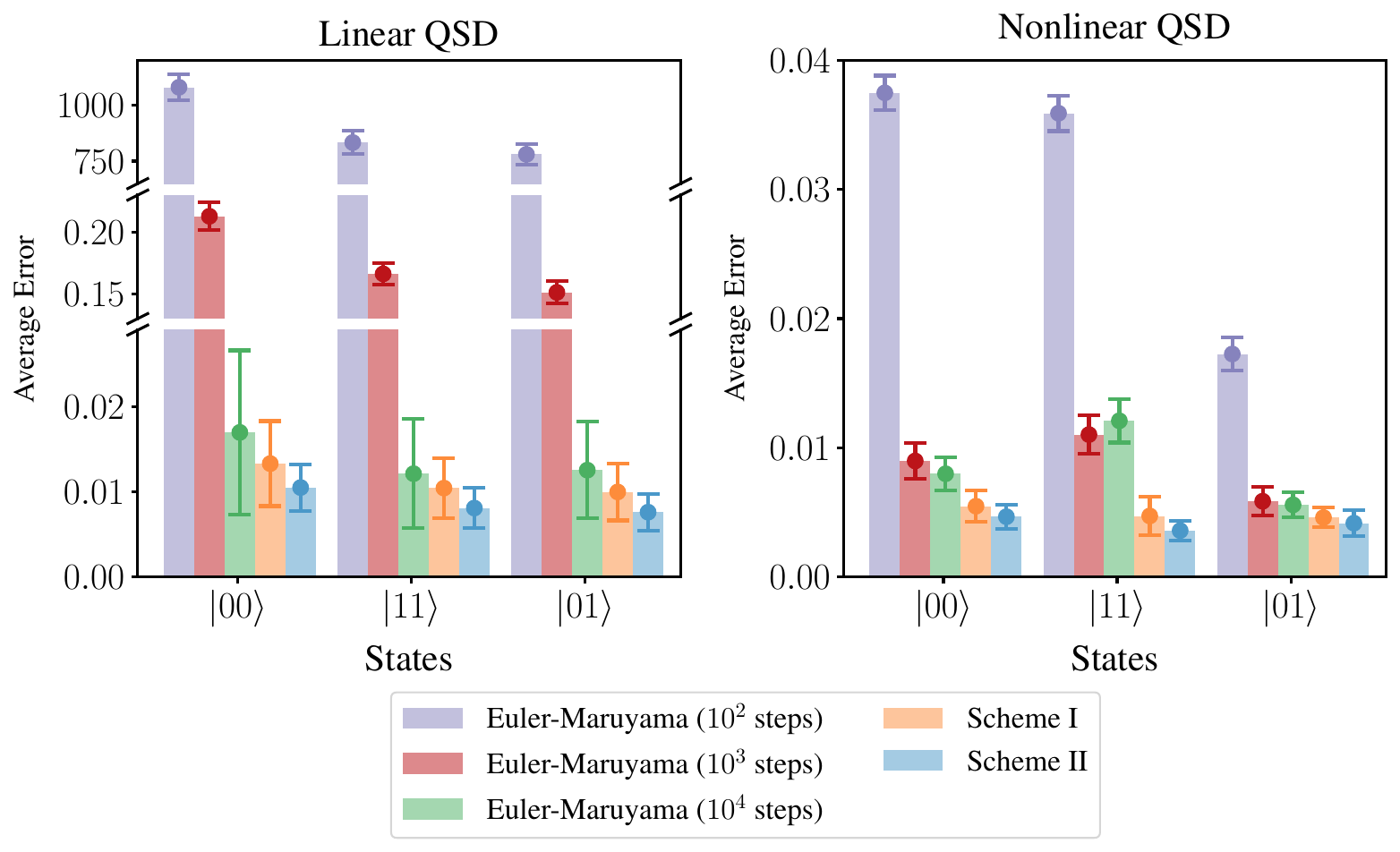}
\caption{\label{fig:comparison_EM_Magnus} \raggedright The comparison of the performance between Euler-Maruyama method (three bars on the left) and the Magnus integrator method (two bars on the right) for solving the dynamics of TFIM with damping. We set the step length $\Delta = 2.5 \times 10^{-3}$, $2.5 \times 10^{-2}$ and $2.5 \times 10^{-1} tJ$ for Euler-Maruyama and $\Delta = 2.5 \times 10^{-1} tJ$ for the Magnus integrator. We repeat the experiment for $10$ times and evaluate the average error of the
populations at each step throughout the simulation. The error bars represent the $99\%$ confidence interval.}
\end{figure}

The results are showcased in \cref{fig:comparison_EM_Magnus}. We observe that in the linear unraveling, the Euler–Maruyama method even fails to produce meaningful results when using time steps of $\Delta = 2.5\times 10^{-1}~tJ$ or $2.5\times 10^{-2}~tJ$. Even at a much smaller time step of $\Delta = 2.5\times 10^{-3}~tJ$, its accuracy remains substantially inferior to that of the Scheme I Magnus integrator at a large step size of $\Delta = 2.5\times 10^{-1}~tJ$. A similar trend is observed for nonlinear unraveling: both Scheme I and Scheme II Magnus integrators, using a large step size of $\Delta=2.5\times 10^{-1}~tJ$,  yield significantly better results than the Euler–Maruyama method, even though the latter is carried out with smaller by orders of magnitude step size of $\Delta = 2.5\times 10^{-3}~tJ$.

In classical exact wavefunction treatment, the overhead introduced in the Magnus expansion compared to the Euler-Maruyama method mainly lies in evaluating the action of the matrix exponential on a state vector at each propagation step. However, this operation can be performed efficiently using Krylov subspace methods \cite{LiLi2019}. In the context of quantum implementations, the advantages of the Magnus integrator are even more pronounced, in the sense that unlike the Euler–Maruyama method, the stochastic Magnus approach preserves an exponential integrator structure and admits a variational formulation via the McLachlan projection principle \cref{eq:mclachlan}, making it more suitable for implementation on quantum devices.

\section{Parameter sets used in the numerical experiments}\label{appF}

In this section, we present the parameter values used for the numerical tests in \cref{sec:numerics}. We adopt the parameter settings from \citet{HuHeadMarsdenMazziottiEtAl2022} for FMO model (\cref{sec:fmo_numer}) and from \citet{PooniaKondabagilSahaGanguly2017} for RPM (\cref{sec:rpm_numer}). All the parameter values are summarized in \cref{table:paravalue}.

\begin{table}[h]
    \centering
    \caption{Parameter value used in the numerical experiments}\label{table:paravalue}
    \begin{tabular}{cll}
        \toprule
       
       \multicolumn{1}{c} {\textbf{symbol}} &  \multicolumn{1}{c}{\textbf{description}} &  \multicolumn{1}{c}{\textbf{values}} \\
        \midrule
         $\alpha $ & dephasing rate in FMO & $3\times 10^{-3}$~fs\\
          $\beta $ & dissipation rate in FMO& $5\times 10^{-7}$~fs\\
           $\gamma$ & sink in FMO& $6.28\times 10^{-3}$~fs\\
        $a_x$ & $x$-component of the hyperfine tensor in RPM & $0.345$ G \\
         $a_y$ & $y$-component of the hyperfine tensor in RPM& $0.345$ G \\
          $a_z$ & $z$-component of the hyperfine tensor in RPM& $9$ G \\
    
        $\hbar$ & reduced Planck constant & $1.05457 \times 10^{-34}$ J$\cdot$s \\
        $\mu_B$ &Bohr magneton & $9.27401\times 10^{-21}$~G\\
        $g$ & the electron-spin $g$-factor used in RPM& $2$\\
        $B_0$  & the magnetic induction intensity of the geomagnetic field &  $0.47$ G\\
        $\phi$ & angle between the $x$-axis of the radical pair and the magnetic field & 0 \\
        \bottomrule
    \end{tabular}
\end{table}
\end{document}